\documentclass[a4paper,12pt]{article}
\pdfoutput=1
\usepackage{amssymb}
\usepackage{tipa}
\usepackage{mathrsfs}
\usepackage{bbm}
\usepackage{epsf,amsmath}
\usepackage[dvipsnames,usenames]{color}
\usepackage{graphicx,subfigure}
\usepackage{color}
\usepackage{colortbl}
\usepackage{hyperref}

\renewcommand{\baselinestretch}{1.5}
\newlength{\dinwidth}
\newlength{\dinmargin}
\def\be{\begin{equation}}
\def\ee{\end{equation}}
\def\ba{\begin{eqnarray}}
\def\ea{\end{eqnarray}}
\newcommand{\half}{\frac{1}{2}}

\begin{document}

\title{\bf A Comprehensive Analysis of Hadronic $b\to s$ Transitions in a Family Non-universal $Z^{\prime}$ Model}
\bigskip
\author{Qin Chang$^{a,b}$\footnote{changqin@htu.edu.cn}, Xin-Qiang Li$^{b,c}$\footnote{xqli@itp.ac.cn} and Ya-Dong Yang$^{b}$\footnote{yangyd@iopp.ccnu.edu.cn}\\
{ $^a$\small Institute of Particle and Nuclear Physics, Henan Normal University, Henan 453007, P.~R. China}\\
{ $^b$\small Institute of Particle Physics and Key Laboratory of Quark and Lepton Physics~(MOE),}\\[-0.2cm]
{     \small Central China Normal University, Wuhan, Hubei 430079, P.~R. China}\\
{ $^c$\small State Key Laboratory of Theoretical Physics, Institute of Theoretical Physics,}\\[-0.2cm]
{     \small Chinese Academy of Sciences, P.~R. China}}

\date{}
\maketitle

\begin{abstract}
{\renewcommand\baselinestretch{1}\selectfont \noindent
Motivated by the latest improved measurements of B-meson decays, we make a comprehensive analysis of the impact of a family non-universal $Z^{\prime}$ boson on $B_s-\bar{B}_s$ mixing and two-body hadronic B-meson decays, all being characterized by the quark-level $b\to s$ transition. Explicitly 22 decay modes and the related 52 observables are considered, and some interesting correlations between them are also carefully examined. Firstly, the allowed oases of $b-s-Z^{\prime}$ coupling parameters $|B^{L,R}_{sb}|$ and $\phi^{L,R}_s$ are extracted from $B_s-\bar{B}_s$ mixing. Then, in the ``SM limit"~({\it i.e.,} no new types of $Z^{\prime}$-induced four-quark operators arise compared to the SM case), we study the $Z^{\prime}$ effects on $B\to\pi K$, $\pi K^{\ast}$ and $\rho K$ decays. It is found that a new weak phase $\phi^{L}_s\sim -90^{\circ}$ is crucial for resolving the observed ``$\pi K$ CP puzzle'' and the allowed oases of the other $Z^{\prime}$ coupling parameters are also strongly restricted. Moreover, the $Z^{\prime}$ effects on $\bar{B}_s\to K K$, $K K^{\ast}$ and $\pi^0 \phi$ decays, being induced by the same quark-level $b\to s q\bar{q}~(q=u,d)$ transitions, are also investigated. Especially, it is found that the decay $\bar{B}_s \to \pi^0 \phi$, once measured, would play a key role in revealing the observed ``$\pi K$ CP puzzle" and probing possible new physics hints. Finally, to check the non-universality of $Z^{\prime}$ couplings to light-quark pairs, we have studied the $B\to \phi K$ decays in detail and found that the left-handed $s-s-Z^{\prime}$ coupling is different from the $d-d-Z^{\prime}$ one, which is due to the large $A_{CP}^{dir}(B^-\to\phi K^-)$ reported by the BaBar collaboration.
\par}
\end{abstract}

\noindent{{\bf Keywords:} B-Physics; Rare Decays; Beyond Standard Model; CP violation}

\newpage

\section{Introduction}

With the fruitful running of BaBar, Belle, Tevatron, LHCb and the coming Super-KEKB experiments, rare B-meson decays play a vital role in precisely testing the Standard Model~(SM) and deciphering the flavour structure of possible New Physics~(NP) models. Although most of the experimental measurements are in good agreement with the SM predictions, some tensions or the so-called puzzles have been observed in the quark-flavour sector~\cite{HFAG} over the past few years. With more statistics collected, some of them have gone, but some still persist and are confirmed by independent measurements.

For instance, for the CP-violating phase $\phi_s^{c\bar{c}s}$, which is defined as the weak phase difference between the $B_s-\bar{B}_s$ mixing amplitude and the $b\to c\bar{c}s$ decay amplitude, a large deviation from the SM prediction was observed by the CDF~\cite{CDFphiold} and D0~\cite{D0phiold} collaborations around 2008. A combined model-independent analysis performed by the UTfit collaboration found that the discrepancy was even more than $3\sigma$~\cite{UTfit}, which attracted much attention. In our previous paper~\cite{changsllzp}, we pursued possible solution through a family non-universal $Z^{\prime}$ boson and constrained the parameter space with these old data. However, the most recent updated measurements from CDF~\cite{CDFDG}, D0~\cite{D0DG}, ATLAS~\cite{ATLASDG} and LHCb~\cite{LHCbDG} are now in good agreement with the SM expectation. Furthermore, beside the CP-violating phase $\phi_s^{c\bar{c}s}$, measurements of the other observables related to $B_s-\bar{B}_s$ mixing, including the mass difference $\Delta M_s$, the decay width difference $\Delta\Gamma_s$ and the like-sign dimuon charge asymmetry, have also been updated recently, which could put a much stronger constraint on various NP models. Therefore, it is worth to reinvestigate the NP effects with these updated experimental data.

As is known, the four $B\to \pi K$ decays are among the most important hadronic B-meson decay modes with rich phenomenology. While their branching fractions have all been measured with high precision, it is still very difficult to explain the so-called ``$\pi K$ CP puzzle", {\it i.e.,} why does the difference between the measured direct CP asymmetries $A_{CP}(B^-\to\pi^{0}K^{-})$ and $A_{CP}(\bar{B}^0\to\pi^{+} K^{-})$ differ from zero by $\sim 5.7\sigma$. It is noted that a family non-universal $Z^{\prime}$ model could provide a possible solution to the observed ``$\pi K$ CP puzzle"~\cite{Barger,changpikzp}. Since many other decay modes, such as $B\to\pi K$, $\pi K^{\ast}$ and $\rho K$, as well as $B_s\to K K^{(\ast)}$ and $B_{s}\to\pi^0\phi$, involve also the same quark-level $b\to s q\bar{q}$~($q=u,d$) transitions, a combined investigation with these closely related decay modes taken into account at the same time is very necessary. Moreover, the correlations between observables of these decays are powerful probes of NP effects.

In a family non-universal $Z^{\prime}$ model, the $Z^{\prime}$ couplings to different quarks are generally different from each other. Focusing on the hadronic B-meson decays induced by quark-level $b\to s$ transitions, one may check if the flavour-conserving $s-s-Z^{\prime}$ coupling differs from the $d-d-Z^{\prime}$ one. This could be done by studying the penguin-dominated $B\to \phi K$ decays. With the $b-s-Z^{\prime}$ coupling restricted by $B_s-\bar{B}_s$ mixing, the $s-s-Z^{\prime}$ coupling is then accessible from these decays.

The direct search for $Z^{\prime}$ bosons is also an important physics program of current and future high-energy colliders~\cite{ATLAS-Exotics,CMS-Exotics,Godfrey:2013eta,Gershtein:2013iqa,Kapukchyan:2013hfa}. Present limits from direct production at the LHC and virtual effects at LEP, through interference or mixing with the SM $Z$ boson, imply that the new $Z^{\prime}$ bosons are rather heavy and mix very little with the $Z$ boson. Depending on the considered theoretical model, $Z^{\prime}$ masses of the order of $2.5-3.0~{\rm TeV}$~\cite{ATLAS-Exotics,CMS-Exotics,Aad:2014cka,Chatrchyan:2012oaa} and $Z-Z^{\prime}$ mixing angles at the level of a few per mil~\cite{Erler:2009jh,Andreev:2014fwa} are already excluded. It is expected that a $Z^{\prime}$ boson, if higher than about $5~{\rm TeV}$ and with order one couplings to SM fermions, could be discovered at a high luminosity $\sqrt{s}=14~{\rm TeV}$ LHC run~\cite{Godfrey:2013eta}. The future $e^+e^-$ International Linear Collider~(ILC) with high center-of-mass energies and longitudinally polarized beams could even go beyond the capabilities of the $14~{\rm TeV}$ LHC~\cite{Kapukchyan:2013hfa,Andreev:2012cj,Ananthanarayan:2010bt}. After the discovery of a $Z^{\prime}$ boson at high-energy colliders, further detailed diagnostics of its couplings needs to be done in order to identify the correct theoretical framework.

Motivated by the above arguments, in this paper, we shall perform a comprehensive analysis of the impact of a family non-universal $Z^{\prime}$ boson on hadronic $b\to s$ transitions. Such a family non-universal $Z^{\prime}$ model, featured by tree-level flavour-changing neutral current~(FCNC) and new CP-violating effect beyond the Cabibbo-Kobayashi-Maskawa~(CKM) picture~\cite{ckm}, has been detailed in Refs.~\cite{Langacker1,Langacker2} and attracted much attention in recent years, for instance in Refs.~\cite{Barger,chiang1,Barger1,BZprime,BurasZp}. The study of $Z^{\prime}$ effects on low-energy flavour physics is very important for the direct searches and the specific model building. In our previous works~\cite{changsllzp,changpikzp,changkllzp,changphillzp}, we have performed detailed investigations of the explicit structures of the effective $Z^{\prime}$ chiral coupling matrices and its effects in $b\to s$ transitions. Since then, measurements of many related observables have been significantly refined, which might affect our previous analyses and conclusions. It is, therefore, necessary to make a comprehensive reanalysis of these interesting processes, which are summarized and classified in Table~\ref{clas}. Our strategy is the following: Firstly, we extract the information about flavour-changing $b-s-Z^{\prime}$ coupling from $B_s-\bar{B}_s$ mixing. Then, with the restricted $b-s-Z^{\prime}$ coupling as input, we discuss the flavour-conserving $Z^{\prime}$ couplings through the other decay modes listed in Table~\ref{clas}. Meanwhile, the space of $b-s-Z^{\prime}$ coupling could possibly be further bounded. After that, it is expected to get the explicit numerical results of the effective $Z^{\prime}$ chiral coupling matrices and to check if the $Z^{\prime}$ couplings are universal for the first two generations.

\begin{table}[t]
 \begin{center}
 \caption{\small \label{clas} Summary and classification of the decay modes considered in this paper according to the involved $Z^{\prime}$ couplings.}
 \vspace{0.2cm}
 \footnotesize
 \doublerulesep 0.8pt \tabcolsep 0.35in
 \begin{tabular}{l|l|l} \hline \hline
   Transition                         &Decay modes                         & $Z^{\prime}$ couplings  involved   \\\hline
$|\Delta B|=|\Delta S|=2$                      &$B_s-\bar{B}_s$ mixing        &$b-s-Z^{\prime}$      \\\hline
 $b\to s q\bar{q}~(q=u,d)$ &$B_{u,d}\to\pi K,\, \pi K^{\ast},\,\rho K$   &$b-s-Z^{\prime}$     \\
                                           &$B_s \to K K,\, K K^{\ast},\,\pi^0 \phi$   &   $u-u-Z^{\prime}$, $d-d-Z^{\prime}$   \\\hline
 $b\to s s\bar{s}$               &$B_{u,d}\to \phi K$ &$b-s-Z^{\prime}$,  $s-s-Z^{\prime}$    \\
 \hline \hline
 \end{tabular}
 \end{center}
 \end{table}

Our paper is organized as following. In section~2, we give a brief overview of the family non-universal $Z^{\prime}$ model. In sections~3 and 4, its effects on the $B_s-\bar{B}_s$ mixing and the hadronic $b\to s$ transitions are discussed, respectively. We conclude in section~5. The relevant input parameters are collected in the Appendix.

\section{Overview of the family non-universal $Z^{\prime}$ model}

In several well-motivated extensions of the SM, such as certain string constructions~\cite{string}, $E_6$ models~\cite{E6}, and theories with large extra dimensions, an additional $U(1)^{\prime}$ gauge symmetry and the associated $Z^{\prime}$ gauge boson could arise. Due to the non-diagonal chiral coupling matrix in the mass eigenstate basis, such a new $Z^{\prime}$ boson could have family non-universal couplings to the SM fermions and lead to FCNC processes even at the tree level, which are strictly forbidden within the SM. The basic formalism has been detailed in Refs.~\cite{Langacker1,Langacker2,chiang1} in a way independent of the specific $Z^{\prime}$ model. For consistence and convenience, we shall recapitulate it below.

For a general NP model extended by an extra $U(1)^{\prime}$ gauge symmetry, the $Z^{\prime}$ part of the neutral-current Lagrangian in the gauge-eigenstate basis can be written as~\cite{Langacker1}
\begin{equation}\label{LZp}
\mathcal {L}=-g^{\prime}J_{\mu}^{\prime}Z^{\prime\mu}\,,
\end{equation}
where $g^{\prime}$ is the gauge coupling constant associated with the extra $U^{\prime}(1)$ group at the electro-weak~(EW) scale, and $J_{\mu}^{\prime}$ is the $Z^{\prime}$-induced neutral chiral current given by
\begin{equation}
J_{\mu}^{\prime}=\bar{\psi}_i\gamma_{\mu}\Big[\epsilon^{\psi_L}_{ij}\, \frac{1-\gamma_5}{2}\, + \,\epsilon^{\psi_R}_{ij}\, \frac{1+\gamma_5}{2}\Big]\psi_j\,,
\end{equation}
with $\psi$ being the chiral field of fermions and $i, j$ the family indices. If the diagonal $U^{\prime}(1)$ chiral charge is non-universal for different families, nonzero flavour-changing $Z^{\prime}$ couplings could be generated through fermion mixing. After diagonalizing the Yukawa couplings to quarks by the unitary matrices $V_{\psi_{L,R}}$~(which give the CKM matrix $V_{\rm CKM}=V_{u_L}V_{d_L}^{\dag}$), the $3\times3$ $Z^{\prime}$ chiral coupling  matrices $\epsilon^{\psi_X}$ can be rewritten as
\begin{equation}\label{3}
 B^{\psi_X}=V_{\psi_X}\epsilon^{\psi_X}V_{\psi_X}^{\dagger}\,,\qquad ( X=L,R)
\end{equation}
in the quark mass-eigenstate basis. Here the off-diagonal elements, at least for $B^{\psi_X}_{13}$ and $B^{\psi_X}_{23}$, are generally complex parameters, while the diagonal ones are real due to the hermiticity of the Lagrangian. To be consistent with the convention for the effective Hamiltonian given by Eqs.~(\ref{Heff}) and (\ref{heffz2}), we have absorbed into the effective couplings $B^{\psi_X}$ a global factor $(g^{\prime}M_Z)/(g_1M_{Z^{\prime}})$ that always associates with the $Z^{\prime}$ couplings, where $g_1=e/(\sin{\theta_W}\cos{\theta_W})$ and $M_{Z^{\prime}}$ denotes the mass of the new $Z^{\prime}$ gauge boson.

It is known that the most promising channels for $Z^{\prime}$ searches at the hadron colliders are dilepton and dijet final states~\cite{Godfrey:2013eta}. If the $Z^{\prime}$ couplings to leptons are negligible, the overwhelming constraints from LHC resonant searches in dilation final states would be not so informative. Furthermore, the current lower limits on $M_{Z^{\prime}}$ set by the ATLAS and CMS collaborations are obtained only within several benchmark scenarios, assuming the $Z^{\prime}$ couplings to fermions to be of order one~\cite{ATLAS-Exotics,CMS-Exotics,Aad:2014cka,Chatrchyan:2012oaa}. In a most general $Z^{\prime}$ model, on the other hand, a lighter $Z^{\prime}$ boson, being of the order of EW scale, could still be allowed. In this paper, since only the effective couplings $B^{\psi_X}$ are involved and the $Z^{\prime}$ mass is not treated as an independent parameter, we shall assume that these high-energy constraints on $Z^{\prime}$ properties are satisfied.

Starting with the Lagrangian Eq.~(\ref{LZp}) and after integrating out the heavy $Z^\prime$ gauge boson, one can easily obtain the resulting effective $|\Delta B|=1$ and $|\Delta B|=2$ four-fermion interactions induced by tree-level $Z^\prime$ exchange~\cite{Langacker2,chiang1}, which will be presented in the following sections.

\section{$B_s-\bar B_s$ mixing}

In this section, we shall firstly recapitulate the theoretical framework and the current experimental status of $B_s-\bar B_s$ mixing, and then present our numerical results and discussions.

\subsection{Theoretical framework for $B_s-\bar B_s$ mixing}

The $B_s-\bar{B}_s$ mixing is described by the following Schr\"odinger equation
\begin{equation}
i\frac{d}{dt}\left(\begin{array}{c}|B_s(t)\rangle\\|\bar{B}_s(t)\rangle\end{array}\right) =\left(M^s-\frac{i}{2}\Gamma^s\right)\left(\begin{array}{c}|B_s(t)\rangle\\|\bar{B}_s(t)\rangle\, \end{array} \right)\,,
\end{equation}
where $M^s$ and $\Gamma^s$ denote the mass and the decay matrix, respectively. The mass and the width difference between the two mass eigenstates $|B_{H}\rangle$ and $|B_{L}\rangle$ are obtained after diagonalizing $M^s$ and $\Gamma^s$ and are defined, respectively, as~\cite{Lenz1}
\begin{align}\label{eq:deltams-deltagammas-def}
\Delta M_s &\equiv M_{H}-M_{L}=2 |M_{12}^s|\,, \nonumber \\[0.2cm]
\Delta \Gamma_s &\equiv \Gamma_{L}-\Gamma_{H}=2 |\Gamma_{12}^s|\cos\phi_s\,,
\end{align}
where $\phi_s \equiv \arg(-M_{12}^s/\Gamma_{12}^s)$ is the CP-violating phase, with $M_{12}^s$ and $\Gamma^s_{12}$ the off-diagonal elements of the mass and the decay matrix, respectively.

There are another two interesting observables for $B_s-\bar B_s$ mixing, the flavour-specific CP asymmetry $a_{sl}^s$ and the CP-violating phase $\phi_s^{c\bar{c}s}$, which are defined, respectively, as~\cite{Lenz1}
\begin{equation}
a_{sl}^{s}={\rm Im}\frac{\Gamma^s_{12}}{M_{12}^s}=\frac{\Delta M_s}{\Delta \Gamma_s}\tan\phi_s\,, \qquad
\phi_s^{c\bar{c}s}={\rm arg}(M_{12}^s)\,.
\end{equation}
The CP-violating phase $\phi_s^{c\bar{c}s}$ appears in tree-dominated $b\to c\bar{c}s$ $B_s$ decays like $B_s\to J/\psi\phi$ and $B_s\to J/\psi \pi^+\pi^-$, taking possible mixing effect into account. It should be noted that $\phi_s^{c\bar{c}s}\neq\phi_s$ unless the terms proportional to $V_{cb}V_{cs}^{\ast}V_{ub}V_{us}^{\ast}$ and $(V_{ub}V_{us}^{\ast})^2$ in $\Gamma^s_{12}$ are neglected~\cite{Lenz1}.

Thus, in order to predict the mixing observables $\Delta M_s$, $\Delta \Gamma_s$, $\phi_s^{c\bar{c}s}$, as well as $a_{sl}^{s}$, we need to know the off-diagonal elements $M_{12}^s$ and $\Gamma_{12}^s$ both within the SM and in the $Z^\prime$ model. The off-diagonal element $M_{12}^s$ can be obtained from
\begin{equation}
2m_{B_s}M_{12}^s=\langle B_s|{\cal H}_{\rm eff}^{\rm full}|\bar{B}_s\rangle\,,
\end{equation}
where the full effective Hamiltonian responsible for $|\triangle B|=2$ transition, including both the SM and $Z^{\prime}$ contributions, can be written as
\begin{equation}\label{Heff}
{\cal H}_{\rm eff}^{\rm full} = \frac{G_F^2}{16\pi^2}\, m_W^2\, (V_{tb}V^{\ast}_{ts})^2\, \Big[C_V^{LL}O_V^{LL} +C_V^{RR}O_V^{RR} +C_V^{LR}O_V^{LR}+C_S^{LR}O_S^{LR}\Big]+{\rm h.c.}\,,
\end{equation}
with the four-quark operators defined as
\ba\label{oper}
&&O_V^{LL}=(\bar{s}b)_{V-A}(\bar{s}b)_{V-A}\,,\qquad O_V^{RR}=(\bar{s}b)_{V+A}(\bar{s}b)_{V+A}\,,\nonumber\\
&&O_V^{LR}=(\bar{s}b)_{V-A}(\bar{s}b)_{V+A}\,,\qquad O_S^{LR}=(\bar{s}b)_{S-P}(\bar{s}b)_{S+P}\,.
\ea
While only $O_V^{LL}$ contributes in the SM~\cite{Buchalla:1996vs}, the first three operators $O_V^{LL}$, $O_V^{RR}$ and $O_V^{LR}$ are all present in the $Z^{\prime}$ model, due to the simultaneous presence of left- and right-handed $Z^{\prime}$ couplings. Moreover, the last operator $O_S^{LR}$ arises through renormalization group evolution~(RGE)~\cite{BurasDF2}. The hadronic matrix elements of these operators can be parameterized as~\cite{BagPara}
\ba
\langle O_V^{LL}\rangle&=&\langle O_V^{RR}\rangle=\frac{8}{3}m_{B_s}^2f_{B_s}^2B_1(\mu_b)\,,\nonumber\\
\langle O_V^{LR}\rangle&=&-\frac{4}{3}\Big(\frac{m_{B_s}}{m_b(\mu_b)+m_s(\mu_b)}\Big)^2 m_{B_s}^2f_{B_s}^2B_5(\mu_b)\,,\nonumber\\
\langle O_S^{LR}\rangle&=&2\Big(\frac{m_{B_s}}{m_b(\mu_b)+m_s(\mu_b)}\Big)^2m_{B_s}^2f_{B_s}^2B_4(\mu_b)\,.
\ea

After neglecting the effects of RGE between the scales $\mu_{Z^{\prime}}$ and $\mu_{W}$ and the $Z-Z^{\prime}$ mixing characterized by a small mixing angle $\theta\sim\mathcal {O}(10^{-3})$~\cite{Erler:2009jh,Andreev:2014fwa,Abreu}, the corresponding Wilson coefficients at the scale $\mu_{W}$ can be written as
\ba\label{11}
C_V^{LL}(\mu_{W})&=&C^{SM}(\mu_{W})+\frac{16\pi^2}{\sqrt{2}G_Fm_W^2}\cdot\frac{B_{sb}^{L}B_{sb}^{L}} {(V_{tb}V^{\ast}_{ts})^2}\,,\\\nonumber
C_V^{RR}(\mu_{W})&=&\frac{16\pi^2}{\sqrt{2}G_Fm_W^2}\cdot\frac{B_{sb}^{R}B_{sb}^{R}} {(V_{tb}V^{\ast}_{ts})^2}\,,\\\nonumber
C_V^{LR}(\mu_{W})&=&\frac{16\pi^2}{\sqrt{2}G_Fm_W^2}\cdot\frac{2B_{sb}^{L}B_{sb}^{R}} {(V_{tb}V^{\ast}_{ts})^2}\,,\\\nonumber
C_S^{LR}(\mu_{W})&=&0\,,
\ea
with the SM contribution given by~\cite{Buchalla:1996vs}
\be
C^{SM}(\mu_{W})=S_0(x_t)+\frac{\alpha_s(\mu_{W})}{4\pi}\big[S_1(x_t)+F(\mu_W)S_0(x_t)+B_t S_0(x_t)\big]\,.
\ee
The $Z^{\prime}$ contributions are encoded by the off-diagonal left- and right-handed $b-s-Z^{\prime}$ couplings $B_{sb}^{L,R}=|B_{sb}^{L,R}|e^{i\phi_s^{L,R}}$, where $\phi_s^{L,R}$ denote the corresponding weak phases. Further RGE of these Wilson coefficients from the scale $\mu_W$ down to $\mu_b$ is the same as in the SM~\cite{BurasDF2}.

Within the SM, the off-diagonal element $\Gamma^s_{12}$ can be written as~\cite{Lenz2}
\begin{flalign}
\Gamma^s_{12}&=-\Big[\lambda_c^2\Gamma^{cc}_{12}+2\lambda_c\lambda_u\Gamma^{uc}_{12}+\lambda_u^2 \Gamma^{uu}_{12}\Big]\nonumber\\[0.2cm]
&=-\Big[\lambda_t^2\Gamma^{cc}_{12}+2\lambda_t\lambda_u(\Gamma^{cc}_{12}-\Gamma^{uc}_{12}) +\lambda_u^2(\Gamma^{cc}_{12}-2\Gamma^{uc}_{12}+\Gamma^{uu}_{12})\Big]\,,
\end{flalign}
with the CKM factors $\lambda_i=V_{ib}V_{is}^{\ast}$ for $i=u,c,t$. The explicit expressions for $\Gamma_{12}^{cc,uu,uc}$ could be found in Refs.~\cite{Lenz2,BenekeGam}. It should be noted that, while the $Z^{\prime}$ correction could significantly affect $M_{12}^{s}$, its effect on $\Gamma_{12}^s$ is numerically negligible~\cite{xqLi}, since $\Gamma_{12}^s$ is dominated by the CKM-favoured $b\to c\bar{c}s$ tree-level part $\Gamma^{cc}_{12}$ within the SM~\cite{Lenz2,BenekeGam}.

\subsection{Experimental status of $B_s-\bar{B}_s$ mixing}

The two complex parameters $M_{12}^s$ and $\Gamma_{12}^s$ can be fully determined by the following four observables: the mass difference $\Delta M_s$, the width difference $\Delta\Gamma_s$, the CP-violating phase $\phi_s^{c\bar{c}s}$, as well as the flavour-specific CP asymmetry $a_{sl}^s$. Thanks to the dedicated experimental efforts, all of these four observables have been measured with much improved precision~\cite{HFAG}.

The mass difference $\Delta M_s$ has been precisely measured by the CDF~\cite{CDFDM} and LHCb~\cite{LHCbDM} collaborations, with the averaged value given by~\cite{HFAG}
\be
\Delta M_s=17.69\pm0.08\,{\rm ps^{-1}}\,.
\ee
The value of the width difference $\Delta \Gamma_s$, averaged over the measurements by CDF~\cite{CDFDG}, D0~\cite{D0DG}, ATLAS~\cite{ATLASDG} and LHCb~\cite{LHCbDG} collaborations, reads~\cite{HFAG}
\be
\Delta \Gamma_s=0.081\pm0.011\,{\rm ps^{-1}}\,.
\ee
These two results are in good agreement with the most recent SM predictions, $\Delta M_s=17.3\pm2.6\,{\rm ps^{-1}}$ and $\Delta \Gamma_s=0.087\pm0.021\,{\rm ps^{-1}}$~\cite{Lenz3}.

The CP-violating phase $\phi_s^{c\bar{c}s}$ has been measured through the analyses of $B_s\to J/\psi\phi$ and $B_s\to J/\psi \pi^+\pi^-$ decays. Around 2008, the CDF~\cite{CDFphiold} and D0~\cite{D0phiold} results indicated a large deviation from the SM prediction $\phi_s^{c\bar{c}s}=-2\beta_s\simeq-0.036$.
A combined model-independent analysis preformed by the UTfit collaboration found that the discrepancy was even more than $3\sigma$~\cite{UTfit}, which attracted much attention at that time. In Ref.~\cite{changsllzp}, for example, we have used these old data to constrain the parameter space of the flavour-changing $Z^{\prime}$ couplings. However, the most recent updated data, $[-0.60,0.12]$~(CDF, $68\%$ CL)~\cite{CDFDG} and $-0.55^{+0.38}_{-0.36}$~(D0)~\cite{D0DG}, show a better agreement with the SM expectation. Furthermore, the ATLAS and LHCb collaborations have presented their updated measurements, $0.22\pm0.41\pm0.10$~(ATLAS)~\cite{ATLASDG}, $0.07\pm0.09\pm0.01$ and $0.01\pm0.07\pm0.01$~\cite{LHCbDG}~(LHCb), which are also consistent with the SM expectation. Averaging over the up-to-date measurements, the Heavy Flavour Averaging Group~(HFAG) gives~\cite{HFAG}
\be
\phi_s^{c\bar{c}s}=0.04^{+0.10}_{-0.13}\,,
\ee
which now agrees with the SM prediction within $1\sigma$, and is expected to put more stringent constraints on various NP parameter space.

The flavour-specific CP asymmetry $a_{sl}^{s}$ is another important quantity to explore the CP violation in $B_s$ system, and has also been measured by several experiments. With an integrated luminosity of $9.1~{\rm fb}^{-1}$, the D0 collaboration measured the like-sign dimuon charge asymmetry, $A_{sl}^{b}=(-0.787\pm0.172 (stat)\pm0.093 (syst))\%$~\cite{D0dimu1}, which can be expressed as a linear combination of the $B_d$ and $B_s$ parts, $A_{sl}^{b}=C_d\,a_{sl}^{d}+C_s\,a_{sl}^{s}$. Using the B-factory result for $a_{sl}^{d}$ and evaluating the muon impact parameter, the D0 collaboration extracted the result~\cite{D0dimu1}
\be\label{D0in}
a_{sl}^{s}({\rm D0,dimuon})=(1.81\pm1.06)\%\,,
\ee
which deviates from the SM prediction $a_{sl}^{s}({\rm SM})=(1.9\pm0.3)\times10^{-5}$ by about $1.7\sigma$. In addition, by measuring the charge asymmetry of the tagged $B_s^0\to D_s\mu X$ decays, the D0 collaboration has also performed a direct determination of $a_{sl}^{s}$~\cite{D0dimu2}
\be\label{D0dir}
a_{sl}^{s}({\rm D0, direct})=(-1.12\pm0.74 ({\rm stat})\pm0.17({\rm syst}))\%\,.
\ee
On the other hand, a recent similar measurement of $a_{sl}^{s}$ by the LHCb collaboration reads~\cite{LHCbdimu}
\be\label{LHCbdir}
a_{sl}^{s}({\rm LHCb, direct})=(-0.06\pm0.50({\rm stat})\pm0.36 ({\rm syst}))\%\,.
\ee
Different from the earlier D0 result, the LHCb measurement does not confirm the significant deviation from the SM prediction, although their results are consistent with each other due to the large uncertainties involved. Thus, much refined measurements of $a_{sl}^{s}$ are needed to clarify such a discrepancy in $B_s$ system. Averaging over the direct measurements given by Eqs.~(\ref{D0dir}) and (\ref{LHCbdir}), one obtains
\be
a_{sl}^{s}({\rm direct})=(-0.48\pm0.48)\%\,,
\ee
which will be used in the following numerical analysis.

\subsection{Numerical results and discussions}

With the input parameters collected in the Appendix, we give our SM predictions in the third column of Table~\ref{tabpredmix}. To obtain the theoretical uncertainties, we scan randomly the points in the allowed ranges of the inputs. Our results agree with the ones given in Ref.~\cite{Lenz2}, with a bit of differences induced by different values of input parameters. It is also found that there are no significant deviations from the experimental data listed in the second column of Table~\ref{tabpredmix}. Thus, these observables are expected to put strong constraints on the flavour-changing $Z^{\prime}$ couplings.

Including the $Z^{\prime}$ contributions and with the default values of input parameters, we get the numerical results for the amplitudes:
\ba\label{Amp1}
{\cal A}_V^{LL}(SM+Z^{\prime})\times10^{11} &=& (6.03-0.27i)+2.47e^{i2\phi_s^{L}}|B_{sb}^{L}\times10^{3}|^2\,,\\[0.2cm]
\label{Amp2}
{\cal A}_V^{RR}(Z^{\prime})\times10^{11} &=& 2.47e^{i2\phi_s^{R}}|B_{sb}^{R}\times10^{3}|^2\,,\\[0.2cm]
\label{Amp3}
{\cal A}_V^{LR}(Z^{\prime})\times10^{11} &=& -9.60e^{i(\phi_s^{L}+\phi_s^{R})}|B_{sb}^{L}\times10^{3}|
|B_{sb}^{R}\times10^{3}|\,,\\[0.2cm]
\label{Amp4}
{\cal A}_S^{LR}(Z^{\prime})\times10^{11} &=& -7.75e^{i(\phi_s^{L}+\phi_s^{R})}|B_{sb}^{L}\times10^{3}|
|B_{sb}^{R}\times10^{3}|\,,
\ea
which correspond, respectively, to the four operators listed in Eq.~(\ref{oper}). One can find that the $Z^{\prime}$ contributions are comparable to the SM one when $|B_{sb}^{L}|\sim10^{-3}$ and/or $|B_{sb}^{R}|\sim10^{-3}$. Moreover, even though the operator $O_S^{LR}$ cannot be directly generated by the tree-level $Z^{\prime}$ exchange, its contribution is significant due to the large RGE effect. It is also found that ${\cal A}_V^{LL}(Z^{\prime})$ or ${\cal A}_V^{RR}(Z^{\prime})$ will dominate the $Z^{\prime}$ contributions when $B_{sb}^{L}\gg B_{sb}^{R}$ or $B_{sb}^{L}\ll B_{sb}^{R}$. However, if $B_{sb}^{L}\approx B_{sb}^{R}$, the $Z^{\prime}$ contributions will be dominated by ${\cal A}_V^{LR}(Z^{\prime})$ and ${\cal A}_S^{LR}(Z^{\prime})$. Thus, for simplicity, we shall consider the following two limiting scenarios:
\begin{enumerate}
\item[(1).] Scenario LL with $B_{sb}^{R}=0$ assumed~(the case with $B_{sb}^{L}=0$ and $B_{sb}^{R}$ arbitrary is similar);

\item[(2).] Scenario LR with $B_{sb}^{L}=B_{sb}^{R}$ assumed.
\end{enumerate}
Due to the different signs between Eqs.~(\ref{Amp1})--(\ref{Amp2}) and Eqs.~(\ref{Amp3})--(\ref{Amp4}), the dependence of the amplitudes on the weak phases $\phi_s^{L,R}$ is different from each other in the two scenarios.

\begin{table}[t]
 \begin{center}
 \caption{\label{tabpredmix} \small Numerical results for observables in $B_s-\bar{B}_s$ mixing both within the SM and in the $Z^{\prime}$ model.}
 \vspace{0.2cm}
 \doublerulesep 0.8pt \tabcolsep 0.12in
 \begin{tabular}{lccccccccccc} \hline \hline
                                    &\multicolumn{1}{c}{Exp. data}       & SM &\multicolumn{2}{c}{$Z^{\prime}$~(Scenario~LL)} \\\hline
 $\Delta M_s$          &$17.69\pm0.08$                    &$17.07^{+4.64}_{-3.62}$   &$15.16^{+5.08+1.11}_{-3.11-0.01}$  &$16.27^{+4.50}_{-3.39}$    \\
 $\Delta \Gamma_s$ &$0.081\pm 0.011$   &$0.089^{+0.021}_{-0.027}$  &$0.088^{+0.025+0.000}_{-0.024-0.000}$  &$0.087^{+0.020}_{-0.023}$      \\
 $\phi^{c\bar{c}s}$ &$0.04^{+0.10}_{-0.13}$ &$-0.044^{+0.009}_{-0.009}$    &$-0.045^{+0.004+0.101}_{-0.004-0.090}$   &$-0.135^{+0.012}_{-0.010}$                          \\
 $a_{sl}^s(\%)$        &$-0.48\pm0.48$                       &$(2.85^{+0.87}_{-0.68})\times10^{-3}$  &$(-2.12^{+1.31+55.02}_{-0.61-52.30})\times10^{-3}$ &$-0.050^{+0.013}_{-0.010}$   \\
 \hline \hline
 \end{tabular}
 \end{center}
\end{table}

\begin{figure}[t]
\begin{center}
\subfigure[]{\includegraphics[width=7.5cm]{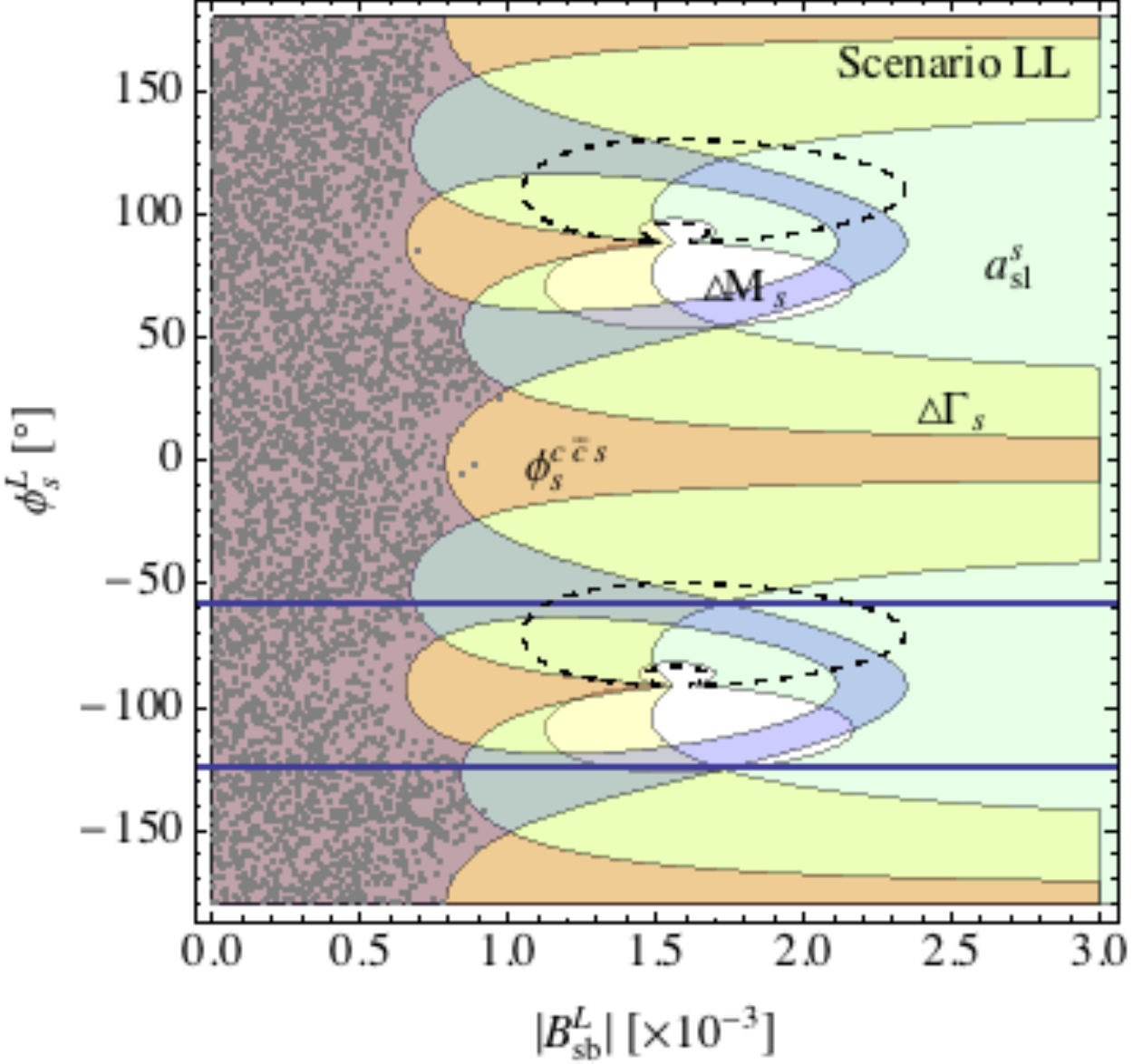}}\qquad
\subfigure[]{\includegraphics[width=7.5cm]{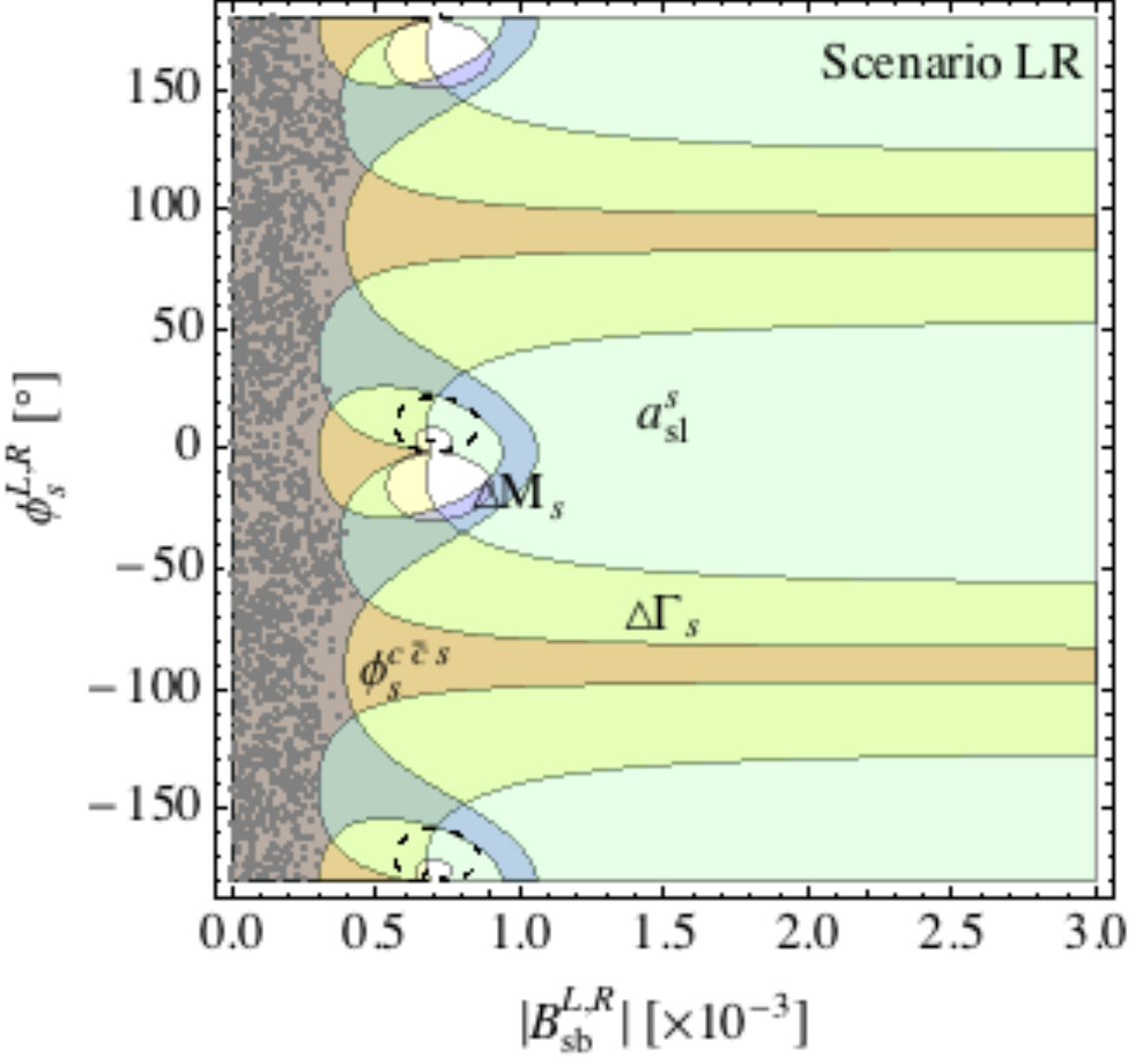}}
\caption{\label{ParaSpac} \small The allowed regions for the parameters $|B_{sb}^{L,R}|$ and $\phi_s^{L,R}$ under the constraints from $\Delta M_s$~(blue), $\Delta \Gamma_s$~(yellow), $\phi_s^{c\bar{c}s}$~(orange), $a_{sl}^{s}$~(light green), as well as their combination~(dotted regions) within $2\sigma$ error bars. The horizontal lines in Fig.~\ref{ParaSpac}(a) correspond to the maximum and minimum values of $\phi_s^{L}$ under the constraints from $B\to\pi K$, $\pi K^{\ast}$ and $\rho K$ decays~(see the text for details).}
\end{center}
\end{figure}

Under the constraints from $\Delta M_s$, $\Delta \Gamma_s$, $\phi_s^{c\bar{c}s}$, $a_{sl}^{s}$, as well as their combination within $2\sigma$ error bars, the allowed parameter spaces for the $Z^{\prime}$ couplings are shown in Fig.~\ref{ParaSpac}. One can see that the precise $\Delta M_s$ and $\phi_s^{c\bar{c}s}$ constrain the $Z^{\prime}$ coupling $B_{sb}^{L}$ strongly, whereas the constraints from $\Delta \Gamma_s$ and $a_{sl}^{s}$ are weak due to the large experimental uncertainties. In both scenarios, the modulus $|B_{sb}^{L}|$ is stringently bounded, and we get numerically
\begin{flalign}
|B_{sb}^{L}|&\leqslant0.98\times10^{-3}\,,\qquad{\rm [Scenario~LL]}\,,\\
|B_{sb}^{L,R}|&\leqslant0.43\times10^{-3}\,,\qquad{\rm [Scenario~LR]}\,.
\end{flalign}
However, no restriction for the weak phases $\phi_s^{L,R}$ is obtained from the updated experimental data, which is quite different from that obtained in  Ref.~\cite{changsllzp}, where the new weak phase $\phi_s^{L}$ is strongly bounded at $\sim-60^{\circ}$ and $\sim-80^{\circ}$ by $B_s-\bar{B}_s$ mixing.

It is noted that the weak phases $\phi_s^{L,R}$ are also restricted by the direct CP violation of hadronic B-meson decays. In the next section as well as in Ref.~\cite{changpikzp}, for Scenario~LL, we find that a weak phase $\phi_s^{L}\sim-90^{\circ}$ is needed to account for the data of $B\to\pi K$, $\pi K^{\ast}$ and $\rho K$ decays, especially the so-called ``$\pi K$ CP puzzle''.  The allowed $\phi_s^{L}$ range $-91^{\circ}\pm33^{\circ}$~(see section 4.2 for detail) is shown as horizontal lines in Fig.~\ref{ParaSpac}(a).  With $\phi_s^{L}$ fixed at this range, we get
\be\label{Zpmix}
|B_{sb}^{L}|\leqslant0.83\times10^{-3}\,,\qquad {\rm with}\,\phi_s^{L}=-91^{\circ}\pm33^{\circ}\,,\qquad {\rm [Scenario~LL]}\,.
\ee
Taking $|B_{sb}^{L}|=0.5\times10^{-3}$ as a benchmark value, which will also be used as the default value in the following sections, we present our final predictions in the fourth column of Table~\ref{tabpredmix}, where the first uncertainty is induced by the SM input parameters, and the second one by the $Z^{\prime}$ parameters $\phi_s^{L}$ given in Eq.~(\ref{Zpmix}). One can find that, with these inputs, all the results agree with the experimental data within uncertainties.

\begin{figure}[t]
\centering
\includegraphics[width=8.5cm]{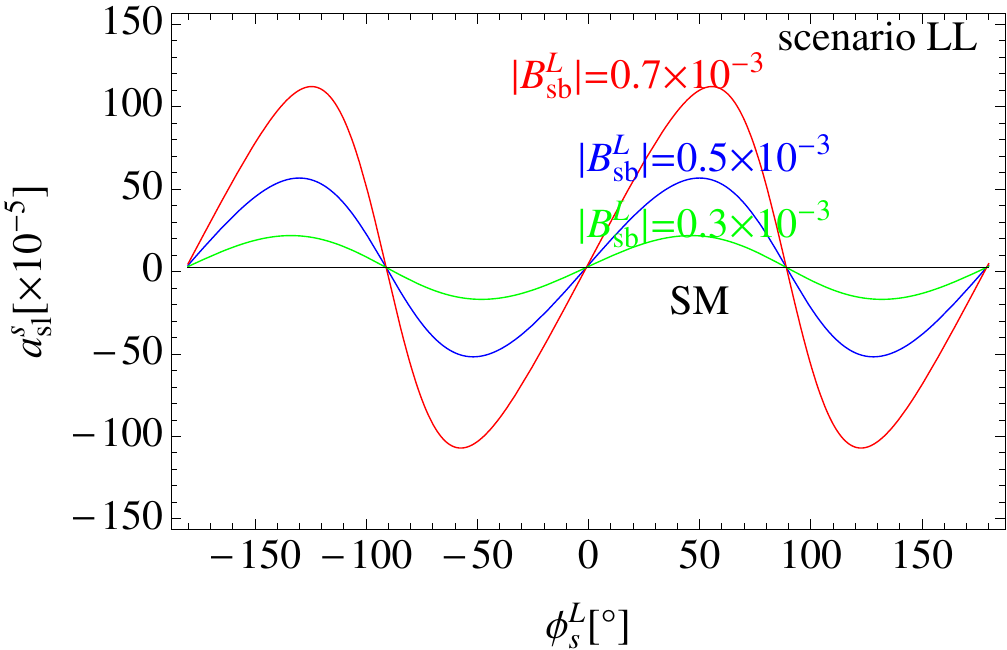}
\caption{\label{aslphis} \small The dependence of $a_{sl}^{s}$ on the weak phase $\phi_s^{L}$, with different values of $|B_{sb}^{L}|$.}
\end{figure}

From Table~\ref{tabpredmix}, it is also found that $a_{sl}^{s}$ is very sensitive to the weak phase $\phi_s^{L}$, and could be either enhanced or reduced by orders of magnitude. The dependence of $a_{sl}^{s}$ on the phase $\phi_s^{L}$, with different values of $|B_{sb}^{L}|$, is shown in Fig.~\ref{aslphis}. It is easily seen that $a_{sl}^{s}$ reaches its minimum value at $\phi_s^{L}=-62^{\circ}$, which is favored for bridging the discrepancy between theoretical prediction and experimental measurement. As a simplified scenario, taking $\phi_s^{L}=-62^{\circ}$ and  $|B_{sb}^{L}|=0.5\times10^{-3}$, we present the corresponding predictions in the last column of Table~\ref{tabpredmix}. One can see that, although being enhanced by 20 times compared to the SM prediction, the predicted $a_{sl}^{s}$ is still one order of magnitude smaller than the central experimental value. Thus, if future refined measurements support $|a_{sl}^{s}|\gg{\cal O}(10^{-4})$, such a family non-universal $Z^{\prime}$ model will suffer a severe challenge. Furthermore, with the indirect result Eq.~(\ref{D0in}) taken into account in the average of $a_{sl}^s$, the constraint from $a_{sl}^{s}$ is shown as black dashed circles in Fig.~\ref{ParaSpac}. One can see that there is no overlap with the combined constraints. So, more accurate experimental measurements are eagerly needed to clarify such a $a_{sl}^{s}$ puzzle observed in $B_s-\bar{B}_s$ mixing.

\section{Hadronic $b\to s q\bar{q}~(q=u,d,s)$ transitions}

With the $b-s-Z^{\prime}$ couplings constrained by $B_s-\bar{B}_s$ mixing, in this section we shall proceed to discuss the impact of $Z^{\prime}$ boson on hadronic $b\to s q\bar{q}~(q=u,d,s)$ transitions.

\subsection{Theoretical framework for $b\to s q\bar{q}~(q=u,d,s)$ transitions}

Within the SM, the effective weak Hamiltonian responsible for the quark-level $b\to s q\bar{q}$ transitions is given by~\cite{Buchalla:1996vs}
\begin{eqnarray}\label{eq:eff}
 {\cal H}_{\rm eff} &=& \frac{G_F}{\sqrt{2}}\,\biggl[V_{ub}
 V_{us}^* \left(C_1 O_1^u + C_2 O_2^u \right) + V_{cb} V_{cs}^* \left(C_1
 O_1^c + C_2 O_2^c \right) - V_{tb} V_{ts}^*\, \big(\sum_{i = 3}^{10}
 C_i O_i \big. \biggl. \nonumber\\[0.2cm]
 && \biggl. \big. + C_{7\gamma} O_{7\gamma} + C_{8g} O_{8g}\big)\biggl] +
 {\rm h.c.},
\end{eqnarray}
where $V_{qb} V_{qs}^*$~($q=u,c,t$) are products of the CKM matrix elements, and $C_{i}$ the Wilson coefficients of the corresponding dimension-six operators.

Starting from the Lagrangian Eq.~(\ref{LZp}) and with the assumption that only the left-handed flavour-changing $Z^{\prime}$ coupling $B_{sb}^L$ is nonzero, the tree-level $Z^{\prime}$-induced effective Hamiltonian for $b\to s q\bar{q}~(q=u,d)$ transitions can be written as
\begin{equation}\label{heffz1}
 {\cal H}_{\rm eff}^{Z^{\prime}} = \frac{2G_F}{\sqrt{2}}\,B_{sb}^L\,(\bar{s}b)_{V-A}\,\sum_{q}\Big[B_{qq}^L (\bar{q}q)_{V-A}+B_{qq}^R(\bar{q}q)_{V+A}\Big]+{\rm h.c.}\,,
\end{equation}
It is noted that, within our assumption, the forms of the above operators already exist within the SM. Thus, in accordance with the SM expression Eq.~(\ref{eq:eff}), we can rewrite Eq.~(\ref{heffz1}) as
\begin{equation}\label{heffz2}
 {\cal H}_{\rm eff}^{Z^{\prime}} = -\frac{G_F}{\sqrt{2}}\,V_{tb}V_{ts}^{\ast}\, \sum_{q}\Big(\Delta C_3 O_3^q +\Delta C_5 O_5^q+\Delta C_7 O_7^q+\Delta C_9
  O_9^q\Big)+{\rm h.c.}\,,
\end{equation}
where $O_i^q~(i=3,5,7,9)$ are the effective four-quark operators, and $\Delta C_i$ the modifications to the corresponding Wilson coefficients due to the tree-level $Z^{\prime}$ exchange. In terms of the model parameters, these Wilson coefficients at the $M_W$ scale are given by
\begin{eqnarray}
 \Delta C_{3,5}&=&-\frac{2}{3V_{ts}^{\ast}V_{tb}}\,B_{sb}^L\,P_{ud}^{L,R}\,, \label{c35} \\
 \Delta C_{9,7}&=&-\frac{4}{3V_{ts}^{\ast}V_{tb}}\,B_{sb}^L\,D_{ud}^{L,R}\,, \label{c79}
\end{eqnarray}
where the off-diagonal coupling $B_{sb}^L=|B_{sb}^L|e^{\phi_s^L}$ is generally complex, and the parameters $P_{ud}^{L,R}$ and $D_{ud}^{L,R}$ are linear combinations of the real flavour-conserving $Z^{\prime}$ couplings and read
\begin{eqnarray} \label{PLR}
P_{ud}^{L,R}&=&B_{uu}^{L,R}+2B_{dd}^{L,R}\,,\\
\label{DLR}
D_{ud}^{L,R}&=&B_{uu}^{L,R}-B_{dd}^{L,R}\,.
\end{eqnarray}
For the $b\to s s\bar{s}$ transition, the $Z^{\prime}$-induced effective Hamiltonian is still given by Eq.~(\ref{heffz2}), but with the corresponding Wilson coefficients modified to
\begin{eqnarray}
 \Delta C_{3,5}&=&-\frac{4}{3V_{ts}^{\ast}V_{tb}}\,B_{sb}^L\,B_{ss}^{L,R}\,, \label{c35s} \\
 \Delta C_{9,7}&=&\frac{4}{3V_{ts}^{\ast}V_{tb}}\,B_{sb}^L\,B_{ss}^{L,R}\,. \label{c79s}
\end{eqnarray}

To evaluate the hadronic matrix elements of $O_i$, we shall adopt the QCD factorization~(QCDF) approach, which has been extensively studied within the SM in Refs.~\cite{Beneke1,Beneke2,Beneke3,Beneke4,Cheng1,Cheng2}. It should, however, be noted that the framework suffers from the endpoint divergences during the evaluation of the hard spectator-scattering and annihilation corrections. The endpoint divergent integrals are usually treated as signs of infrared sensitive contributions and can be parameterized with at least two phenomenological parameters, for example, $X_A=\int^{1}_{0}dy/y=\mathrm{ln}(m_b/\Lambda_h)\,(1+\rho_A e^{i\phi_{A}})$~\cite{Beneke3}. The different scenarios corresponding to different choices of $\rho_{A,H}$ and $\phi_{A,H}$ have been thoroughly discussed in Refs.~\cite{Beneke3,Beneke4,Cheng1,Cheng2}. As an alternative scheme, one could use an infrared-finite gluon propagator, $1/(k^2+i\epsilon)\to1/(k^2-M_g(k^2)+i\epsilon)$~\cite{Cornwall}, to regulate the divergent integrals, which have been thoroughly studied in Refs.~\cite{ChangpikT,changann}. In the latter scheme, it is found that the hard spectator-scattering contributions are real and the annihilation corrections are complex with a large imaginary part~\cite{ChangpikT}. Moreover, the strength of the annihilation corrections is sensitive to the effective gluon mass scale $m_g$, which is the only input parameter with a typical value $0.5\pm0.2~{\rm GeV}$ obtained by relating the gluon mass to the gluon condensate~\cite{Cornwall}. Interestingly, a similar result $m_g=0.5\pm0.05~{\rm GeV}$ is also obtained with the constraints from $B_{u,d}\to\pi K$, $\pi K^{\ast}$ and $\rho K$ decays taken into account~\cite{ChangpikT}. In this paper, we shall use the second scheme and, for simplicity, take $m_g=0.5\pm0.05~{\rm GeV}$ to regulate the encountered endpoint divergences.

\subsection{$B{\to}\pi K$, $\pi K^{\ast}$ and $\rho K$ decays}

One of the well-known anomalies observed in hadronic $b\to s$ transitions is the so-called ``$\pi K$ CP puzzle"~\cite{pikpuz}, {\it i.e.,} the significant discrepancy between experimental data and theoretical prediction for the difference between the direct CP asymmetries, $\Delta A\equiv A_{CP}(B^{-}\to K^{-}\pi^{0}) - A_{CP}(\bar{B}^{0}\to K^{-}\pi^{+})$. Using the up-to-date averaged results, $A_{CP}(B^{-}\to K^{-}\pi^{0})=(4.0\pm2.1)\%$ and $A_{CP}(\bar{B}^{0}\to K^{-}\pi^{+})=(-8.6\pm0.7)\%$~\cite{HFAG}, one gets
\be\label{Dpik}
\Delta A=(12.6\pm2.2)\%\,,
\ee
which differs from zero by about $5.7\sigma$. Within the SM, however, $A_{CP}(B^{-}\to K^{-}\pi^{0})$ and $A_{CP}(\bar{B}^{0}\to K^{-}\pi^{+})$ are expected to be approximately equal to each other~\cite{pikpuz}.

It is noted that a family non-universal $Z^{\prime}$ model, featured by tree-level FCNCs and new CP-violating phases, could provide a possible solution to the observed ``$\pi K$ CP puzzle"~\cite{Barger}. However, since the $B\to\pi K^{\ast}$, $\rho K$ and $B_s\to K K^{(\ast)}$ decays also involve the same quark-level $b\to s q\bar{q}$~($q=u,d$) transitions, it is necessary to take into account all these decay modes at the same time. In Ref.~\cite{changpikzp}, we have studied in detail the $Z^{\prime}$ effect on $B\to\pi K$, $\pi K^{\ast}$ and $\rho K$ decays and obtained the allowed parameter spaces for the involved $Z^{\prime}$ couplings. In this paper, due to a lot of updated experimental measurements, we shall update the fitting results for the $Z^{\prime}$ couplings under the constraints of these decay modes.

Adopting the same conventions as in Ref.~\cite{Beneke3}, we can write the decay amplitudes for the four $B\to\pi K$ decays, respectively, as
\begin{eqnarray}
{\cal A}_{B^-\to\pi^- \bar{K}}
   &=& \sum_{p=u,c}V_{pb}V_{ps}^{\ast} A_{\pi \bar{K}} \Big[
    \delta_{pu}\,\beta_2 + \alpha_4^p - \half\alpha_{4,{\rm
    EW}}^p +\beta_3^p+\beta_{3,{\rm
    EW}}^p\Big]\,, \\[0.2cm]
\label{amp1}
\sqrt2\, {\cal A}_{B^-\to\pi^0 K^-}
   &=& \sum_{p=u,c}V_{pb}V_{ps}^{\ast} \biggl\{ A_{\pi^0 K^-} \Big[
    \delta_{pu}\,(\alpha_1+\beta_2) + \alpha_4^p + \alpha_{4,{\rm
    EW}}^p +\beta_3^p+\beta_{3,{\rm EW}}^p\Big]\nonumber\\
   &&+  A_{ K^- \pi^0}\Big[\delta_{pu}\,\alpha_2+\frac{3}{2}
   \alpha_{3,{\rm EW}}^p\Big]\biggl\}\,,
\label{amp2}
\end{eqnarray}
\begin{eqnarray}
{\cal A}_{\bar{B}^0\to\pi^+ K^-}
   &=& \sum_{p=u,c}V_{pb}V_{ps}^{\ast} A_{\pi^+ K^-} \Big[
    \delta_{pu}\,\alpha_1 + \alpha_4^p + \alpha_{4,{\rm
    EW}}^p +\beta_3^p-\half\beta_{3,{\rm EW}}^p\Big]\,,\\[0.2cm]
\label{amp3}
\sqrt2\, {\cal A}_{\bar{B}^0\to\pi^0 \bar{K}^0}
   &=& \sum_{p=u,c}V_{pb}V_{ps}^{\ast} \biggl\{ A_{\pi^0 \bar{K}^0}
   \Big[-\alpha_4^p + \half\alpha_{4,{\rm
    EW}}^p -\beta_3^p+\half\beta_{3,{\rm EW}}^p\Big]\nonumber\\
   &&+  A_{ \bar{K}^0
   \pi^0}\Big[\delta_{pu}\,\alpha_2+\frac{3}{2}\alpha_{3,{\rm
   EW}}^p\Big]\biggl\}\,,
\label{amp4}
\end{eqnarray}
where the explicit expressions for the coefficients $\alpha_i^p\equiv\alpha_i^p(M_1M_2)$ and $\beta_i^p\equiv\beta_i^p(M_1M_2)$ can be found in Ref.~\cite{Beneke3}. The decay amplitudes for $B\to\pi K^{\ast}$ and $B\to\rho K$ decays could be obtained from the above ones by replacing $(\pi K)\to (\pi K^{\ast})$ and $(\pi K)\to (\rho K)$, respectively.

With the above mentioned theoretical formulae and the input parameters collected in the Appendix, our SM predictions for the branching fractions, direct and mixing-induced CP asymmetries of $B\to\pi K$, $\pi K^{\ast}$ and $\rho K$ decays are summarized in the third column of Tables~\ref{pikbr}, \ref{pikdircp} and \ref{pikmixcp}, respectively. It is found that most of our theoretical predictions are generally consistent with the experimental data. However, as is expected within the SM, $A_{CP}(B^-\to K^{-}\pi^{0})\sim-11.7\%$ is still very close to $A_{CP}(\bar{B}^0\to K^{-}\pi^{+})\sim-14.5\%$, which has already been observed in Ref.~\cite{changpikzp}.

\begin{table}[t]
 \begin{center}
 \caption{\small The CP-averaged branching ratios~(in units of $10^{-6}$) of $B\to\pi K$, $\pi K^{\ast}$ and $\rho K$ decays both within the SM and in the $Z^{\prime}$ model with the two different scenarios. The first and the second theoretical uncertainties shown in the last three columns are due to the variations of the SM parameters listed in Appendix and the effective gluon mass $m_g=0.5\pm0.05~{\rm GeV}$, respectively. The third theoretical uncertainties in the last two columns are due to the $Z^{\prime}$ couplings listed in Table~\ref{ZpCoupValue1}.} \label{pikbr}
 \vspace{0.1cm}
 \doublerulesep 0.8pt \tabcolsep 0.12in
 \begin{tabular}{lcccccccccccc} \hline \hline
 \multicolumn{1}{c}{Decay Mode}       &\multicolumn{1}{c}{Exp.}&\multicolumn{1}{c}{ SM }&\multicolumn{2}{c}{$Z^{\prime}$ model} \\
                                                                 &     data                                 &                                &Scenario I             &Scenario II    \\ \hline
 $B^-\to\pi^-{\bar K}^0$            &$23.79\pm0.75$  &$19.81^{+7.10+2.41}_{-5.62-1.72}$&$19.85^{+7.75+2.41+0.95}_{-5.59-1.72-0.96}$ &$19.82^{+7.34+2.41+0.13}_{-5.13-1.72-0.13}$\\
 $B^-\to\pi^0K^-$                       &$12.94^{+0.52}_{-0.51}$&$10.75^{+3.76+1.23}_{-2.91-0.88}$&$10.36^{+3.63+1.23+1.66}_{-2.91-0.88-1.27}$ &$10.40^{+3.74+1.23+1.37}_{-2.82-0.88-1.31}$ \\
 ${\bar B}^0\to\pi^{+}K^-$       &$19.57^{+0.53}_{-0.52}$&$16.75^{+6.17+2.17}_{-4.79-1.55}$&$16.37^{+6.05+2.18+1.32}_{-4.78-1.56-1.41}$ &$16.63^{+5.98+2.18+0.09}_{-4.58-1.56-0.07}$\\
 ${\bar B}^0\to\pi^0{\bar K}^0$&$9.93\pm0.49$                 &$7.66^{+3.01+1.07}_{-2.32-0.76}$   &$7.70^{+3.25+1.07+1.93}_{-2.12-0.76-1.00}$  &$7.76^{+2.49+1.07+1.22}_{-2.23-0.76-1.03}$  \\
  \hline
 $B^-\to\pi^-{\bar K}^{\ast0}$    &$9.9^{+0.8}_{-0.9}$   &$8.5^{+2.5+2.0}_{-2.2-1.5}$         &$8.5^{+2.9+2.0+0.6}_{-2.1-1.5-0.5}$ &$8.5^{+2.5+2.0+0.4}_{-2.0-1.5-0.4}$  \\
 $B^-\to\pi^0K^{\ast-}$               &$8.2\pm1.8$                 &$5.2^{+2.0+1.0}_{-1.5-0.7}$         &$4.6^{+1.9+1.0+0.6}_{-1.4-0.7-0.6}$  &$4.6^{+1.6+1.0+0.7}_{-1.3-0.8-0.6}$  \\
 ${\bar B}^0\to\pi^{+}K^{\ast-}$&$8.5\pm0.7$                &$7.8^{+3.1+1.8}_{-2.2-1.3}$         &$8.0^{+3.2+1.8+0.5}_{-2.4-1.3-0.5}$  &$7.9^{+2.8+1.8+0.3}_{-2.2-1.3-0.3}$  \\
 ${\bar B}^0\to\pi^0{\bar K}^{\ast0}$&$2.5\pm0.6$         &$3.2^{+1.2+0.9}_{-1.0-0.6}$        &$3.4^{+1.2+0.9+0.9}_{-0.9-0.6-0.7}$  &$3.4^{+1.0+0.9+0.8}_{-0.8-0.6-0.6}$ \\
  \hline
 $B^-\to\rho^-{\bar K}^0$            &$8.0^{+1.5}_{-1.4}$     &$8.3^{+3.1+2.2}_{-2.3-1.6}$      &$8.8^{+3.1+2.3+1.2}_{-2.3-1.6-1.4}$  &$8.4^{+2.5+2.2+0.9}_{-2.1-1.6-0.8}$ \\
 $B^-\to\rho^0K^-$                       &$3.81^{+0.48}_{-0.46}$&$4.10^{+1.61+1.05}_{-1.09-0.75}$&$3.88^{+1.86+1.10+0.94}_{-1.24-0.79-0.89}$  &$3.66^{+1.64+1.06+0.52}_{-1.15-0.75-0.40}$\\
 ${\bar B}^0\to\rho^{+}K^-$       &$7.2\pm0.9$                   &$10.1^{+3.7+2.3}_{-2.7-1.6}$       &$11.4^{+4.9+2.4+1.3}_{-3.0-1.7-2.2}$ &$10.4^{+3.7+2.3+0.9}_{-2.9-1.7-0.8}$\\
 ${\bar B}^0\to\rho^0{\bar K}^0$&$4.7\pm0.7$                   &$5.4^{+1.7+1.2}_{-1.4-0.9}$      &$6.4^{+2.0+1.2+1.4}_{-1.6-0.9-2.0}$  &$5.8^{+1.7+1.2+1.8}_{-1.4-0.9-1.6}$\\
 \hline \hline
 \end{tabular}
 \end{center}
\end{table}

\begin{table}[t]
 \begin{center}
 \caption{\small The direct CP asymmetries~(in units of $10^{-2}$) of $B\to\pi K$, $\pi K^{\ast}$ and $\rho K$ decays. The other captions are the same as in Table~\ref{pikbr}.}
 \label{pikdircp}
 \vspace{0.2cm}
 \doublerulesep 0.8pt \tabcolsep 0.15in
 \begin{tabular}{lcccccccccccc} \hline \hline
 \multicolumn{1}{c}{Decay Mode}       &{Exp.}  &{ SM }                      &\multicolumn{2}{c}{$Z^{\prime}$ model} \\
                                                                 &     data  &                                        &Scenario I                                                &Scenario II    \\ \hline
 $B^-\to\pi^-{\bar K}^0$            &$-1.5\pm1.9$ &$0.3^{+0.2+0.0}_{-0.2-0.0}$&$1.5^{+1.6+0.8+1.6}_{-1.7-0.8-1.2}$ &$2.8^{+0.9+0.5+0.4}_{-1.0-0.5-0.9}$\\
 $B^-\to\pi^0K^-$                       &$4.0\pm2.1$  &$-11.7^{+3.2+0.8}_{-3.5-0.8}$ &$-0.5^{+4.2+1.0+3.3}_{-3.9-1.0-0.5}$ &$-1.0^{+5.1+0.7+1.3}_{-3.7-0.8-1.4}$\\
 ${\bar B}^0\to\pi^{+}K^-$       &$-8.2\pm0.6$ &$-14.5^{+3.7+0.4}_{-3.6-0.3}$ &$-10.9^{+4.1+1.0+4.1}_{-3.9-0.9-2.3}$ &$-11.4^{+4.1+0.7+0.6}_{-3.7-0.6-0.8}$\\
 ${\bar B}^0\to\pi^0{\bar K}^0$&$-1\pm10$    &$0^{+1+0}_{-1-0}$     &$-7^{+4+1+6}_{-6-1-5}$  &$-9^{+4+1+3}_{-5-0-3}$\\
  \hline
 $B^-\to\pi^-{\bar K}^{\ast0}$    &$-3.8\pm4.2$ &$0.3^{+0.2+0}_{-0.2-0}$   &$-4.7^{+4.3+1.7+13.5}_{-3.7-1.6-11.0}$    &$1.0^{+2.8+0+0.3}_{-2.6-1-0.4}$\\
 $B^-\to\pi^0K^{\ast-}$               &$-6\pm24$    &$-48^{+9+2}_{-10-1}$         &$-29^{+15+0+13}_{-12-0-12}$            &$-22^{+15+0+6}_{-12-0-7}$\\
 ${\bar B}^0\to\pi^{+}K^{\ast-}$&$-23\pm6$    &$-55^{+11+4}_{-10-4}$        &$-58^{+10+3+10}_{-8-2-8}$           &$-56^{+9+3+1}_{-8-2-1}$\\
 ${\bar B}^0\to\pi^0{\bar K}^{\ast0}$&$-15\pm13$&$4^{+2+0}_{-2-0}$    &$-28^{+10+2+16}_{-11-2-16}$             &$-36^{+8+3+10}_{-11-1-11}$\\
  \hline
 $B^-\to\rho^-{\bar K}^0$            &$-12\pm17$     &$1^{+0+0}_{-0-0}$ &$0^{+3+1+11}_{-3-1-9}$       &$-1^{+2+1+1}_{-2-1-0}$\\
 $B^-\to\rho^0K^-$                       &$37\pm11$     &$56^{+15+4}_{-14-4}$      &$7^{+20+1+22}_{-23-1-1}$                 &$8^{+20+0+18}_{-22-0-2}$\\
 ${\bar B}^0\to\rho^{+}K^-$       &$20\pm11$    &$40^{+11+2}_{-10-3}$         &$36^{+10+1+3}_{-10-1-6}$                    &$36^{+10+1+1}_{-10-2-1}$\\
 ${\bar B}^0\to\rho^0{\bar K}^0$&$-6\pm20$     &$-2^{+2+1}_{-2-1}$   &$29^{+7+0+6}_{-7-1-15}$              &$26^{+7+1+7}_{-6-1-11}$\\
 \hline \hline
 \end{tabular}
 \end{center}
\end{table}

\begin{table}[t]
 \begin{center}
 \caption{\small The mixing-induced CP asymmetries~(in units of $10^{-2}$) of $\bar{B}^0\to\pi^0K^{0}$ and $\bar{B}^0\to\rho^0K^{0}$ decays. The other captions are the same as in Table~\ref{pikbr}.}
 \label{pikmixcp}
 \vspace{0.2cm}
 \doublerulesep 0.8pt \tabcolsep 0.26in
 \begin{tabular}{lccccccccccc} \hline \hline
{Decay Mode}                         &{Experiment} &{SM}                       &\multicolumn{2}{c}{$Z^{\prime}$ model}\\
                                                 &{data}            &                                       &Scenario I                         &Scenario II\\ \hline
 ${\bar B}^0\to\pi^0K^0$  & $57\pm17$ & $85^{+7+0}_{-10-0}$              & $73^{+11+1+15}_{-14-1-15}$ & $67^{+11+1+4}_{-17-1-3}$\\
 ${\bar B}^0\to\rho^0 K^0$&$54^{+18}_{-21}$ & $76^{+9+0}_{-9-0}$ &$94^{+3+1+3}_{-6-1-30}$    & $91^{+2+0+1}_{-3-0-2}$\\
 \hline \hline
 \end{tabular}
 \end{center}
 \end{table}

As has already been found in Refs.~\cite{Barger,changpikzp}, a family non-universal $Z^{\prime}$ model could provide a solution to the observed ``$\pi K$ CP puzzle", provided that there is a significant correction to the EW-penguin coefficient $\alpha^p_{3,{\rm EW}}(PP)=a_{9}^{p}-a_7^p$ in the amplitude of $B^-\to K^-\pi^0$ decay~(see Eq.~(\ref{amp2})). However, being of a similar amplitude as that of $B^-\to K^-\pi^0$, the $B^-\to K^-\rho^0$ decay will also receive a significant $Z^{\prime}$ correction and, therefore, provide a further constraint on the $Z^{\prime}$ couplings required to reconcile the observed ``$\pi K$ CP puzzle"~\cite{changpikzp}. Under different simplifications for the $Z^{\prime}$ couplings, four different cases have been systematically investigated in Ref.~\cite{changpikzp}. In this paper, we shall pay our attention to the following two scenarios:
\begin{itemize}
\item Scenario~I: without any simplification for the flavour-conserving $Z^{\prime}$ couplings;
\item Scenario~II: assuming that the right-handed flavour-conserving $Z^{\prime}$ couplings vanish.
\end{itemize}

As the decay modes considered in this and the following sections are all related to the combination of flavour-changing and flavour-conserving $Z^{\prime}$ couplings~(see Eqs.~(\ref{c35})--(\ref{c79s})), with the modulus of the flavour-changing $b-s-Z^{\prime}$ coupling constrained by $B_s-\bar{B}_s$ mixing, from now on we shall focus on the constraint on the weak phase $\phi^{L}_s$ and the flavour-conserving $Z^{\prime}$ couplings imposed by these hadronic $b\to s$ transitions. Taking $|B_{sb}^L|=0.5\times 10^{-3}$ as the default value, our fitting for the other $Z^{\prime}$ parameters is performed with the experimental data varying randomly within their $2\sigma$ error bars, while the theoretical uncertainties are obtained by varying the input parameters within the ranges specified in the Appendix. In addition, the uncertainty of $m_g$ is not considered in the fitting for simplicity.

\subsubsection*{Scenario~I: without any simplification for the flavour-conserving $Z^{\prime}$ couplings}

\begin{figure}[t]
\begin{center}
\subfigure[]{\includegraphics[width=7.3cm]{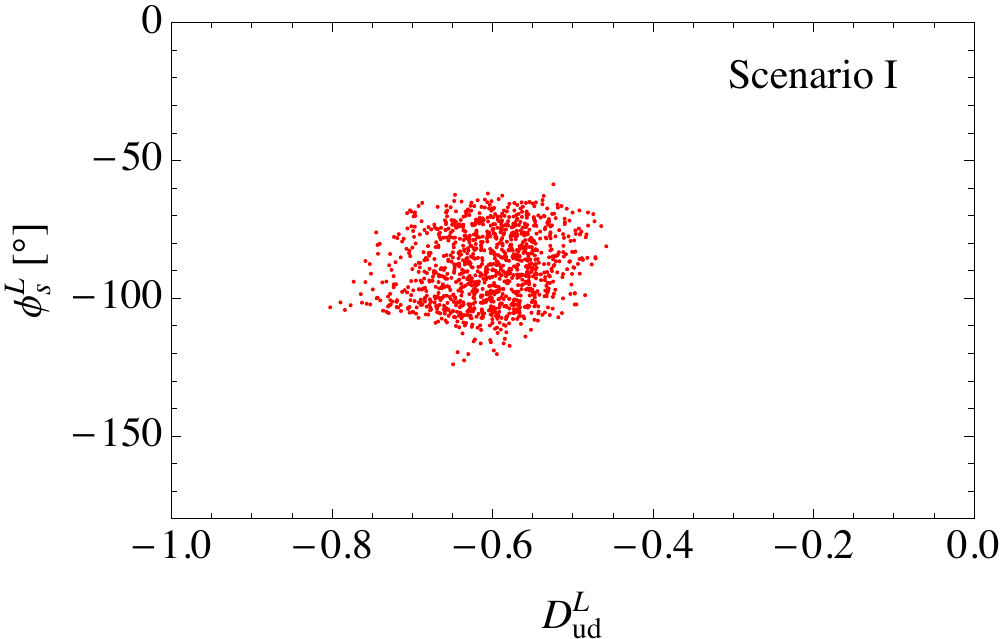}}\qquad
\subfigure[]{\includegraphics[width=7.3cm]{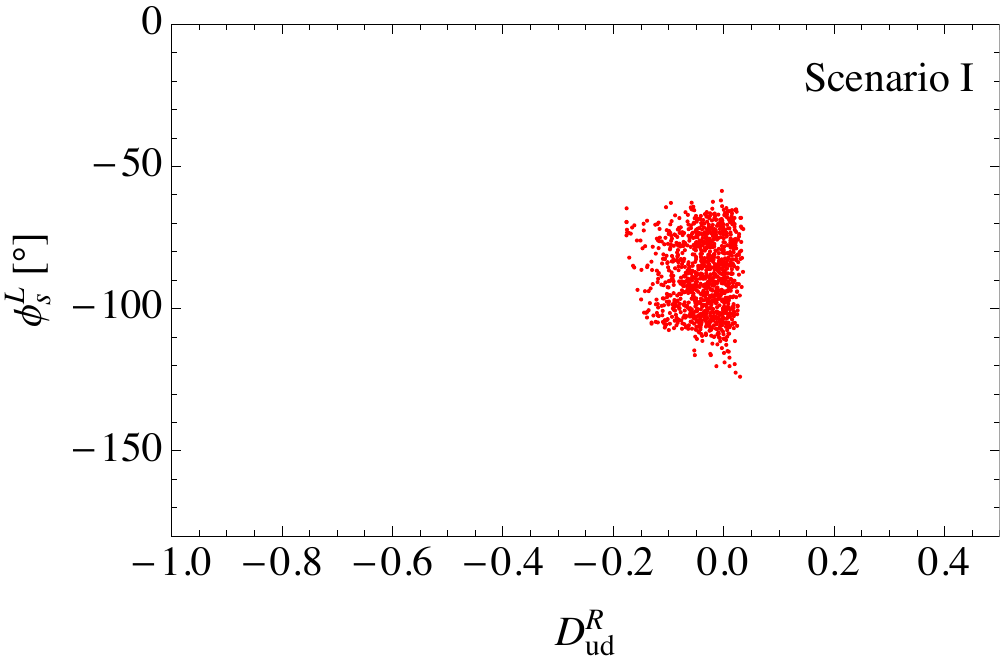}}\\
\subfigure[]{\includegraphics[width=7.3cm]{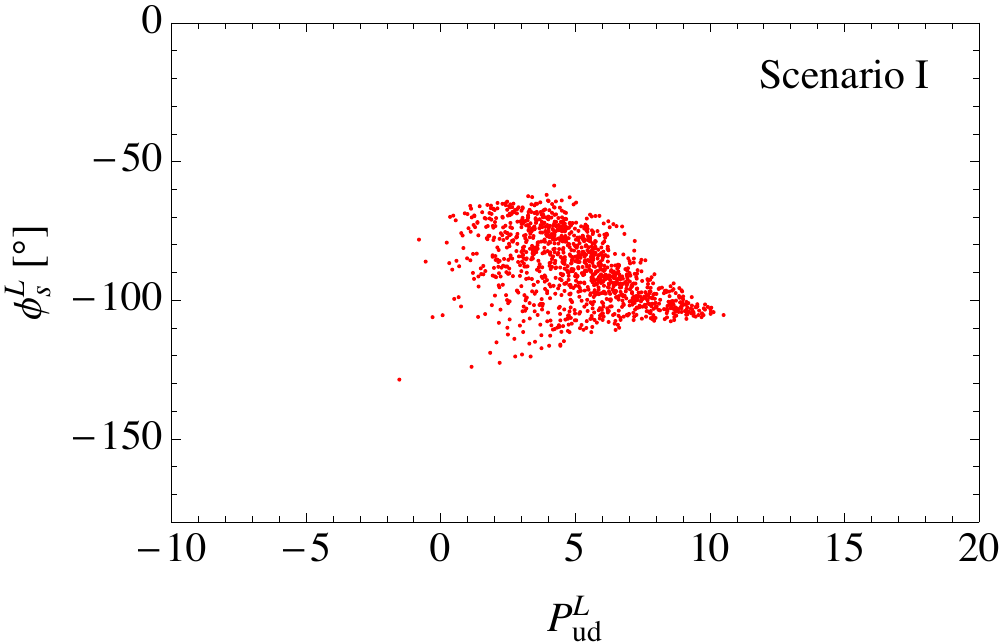}}\qquad
\subfigure[]{\includegraphics[width=7.3cm]{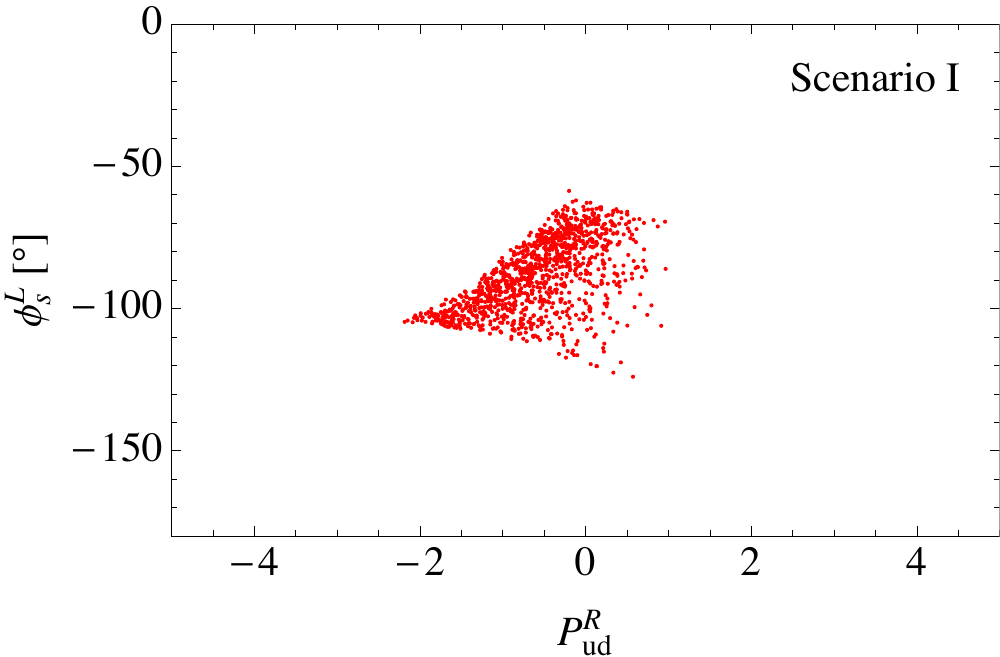}}
\caption{\label{ZpSpac1}\small The allowed regions for the $Z^{\prime}$ coupling parameters in scenario~I, under the constraints from ${\cal B}(B\to \pi K,\rho K)$ and $A_{CP}(B\to \pi K,\rho K)$.}
\end{center}
\end{figure}

As a most general case, we make no simplifications for the flavour-conserving $Z^{\prime}$ couplings in this scenario. Under the constraints from ${\cal B}(B\to \pi K, \rho K)$ and $A_{CP}(B\to \pi K,\rho K)$, the allowed regions for the effective couplings $D_{ud}^{L,R}$ and $P_{ud}^{L,R}$, as well as the weak phase $\phi^{L}_s$ are shown in Fig.~\ref{ZpSpac1}, with the corresponding numerical results given in the second row of Table~\ref{ZpCoupValue1}.

\begin{table}[t]
 \begin{center}
 \caption{\small Numerical results for the flavour-conserving $Z^{\prime}$ couplings $D_{ud}^{L,R}$ and $P_{ud}^{L,R}$, as well as the weak phase $\phi^{L}_s$ in the two different scenarios, in which $|B_{sb}^L|=0.5\times 10^{-3}$ is adopted.}
 \label{ZpCoupValue1}
 \vspace{0.2cm}
 \doublerulesep 0.8pt \tabcolsep 0.13in
 \begin{tabular}{lccccccccccc} \hline \hline
             &$D_{ud}^{L}$ &$D_{ud}^{R}$ &$P_{ud}^{L}$ &$P_{ud}^{R}$ &$\phi^{L}_s[^{\circ}]$\\\hline
Scenario I   &$-0.63\pm0.17$   &$-0.07\pm0.11$ &$4.98\pm5.53$  &$-0.61\pm1.58$  &$-91\pm33$\\
Scenario II &$-0.59\pm0.12$   &---                       &$2.82\pm0.68$  &---                   &$-91\pm31$\\
 \hline \hline
 \end{tabular}
 \end{center}
\end{table}

It is found that the weak phase $\phi^{L}_s$ is bounded to be around $-91^{\circ}$, which has already been used to study the $Z^{\prime}$ effect in $B_s-\bar{B}_s$ mixing, as detailed in section~3. It is also found that such a family non-universal $Z^{\prime}$ model either with a negative $D_{ud}^{L}$ and $\phi^{L}_s\sim -91^{\circ}$ or with a relatively large positive $P_{ud}^{L}$ and $\phi^{L}_s\sim -91^{\circ}$ is required to reconcile the observed ``$\pi K$ CP puzzle". As is shown in Figs.~\ref{ZpSpac1}(b) and \ref{ZpSpac1}(d), on the other hand, the strengths of the right-handed $Z^{\prime}$ couplings $D_{ud}^{R}$ and $P_{ud}^{R}$ could be zero, indicating their effects to be dispensable.

Using the obtained numerical points shown in Fig.~\ref{ZpSpac1} and the relations given by Eqs.~(\ref{PLR}) and (\ref{DLR}), we give also numerical results for the effective $Z^{\prime}$ couplings $B_{uu}^{L,R}$ and $B_{dd}^{L,R}$ in Table~\ref{ZpCoupValue2}. It can be seen that the right-handed couplings are smaller by about one order of magnitude than the left-handed ones. However, since we could only obtain the direct constraints on their combinations $D_{ud}^{L,R}$ and $P_{ud}^{L,R}$, uncertainties of the $Z^{\prime}$ couplings $B_{uu}^{L,R}$ and $B_{dd}^{L,R}$ listed in Table~\ref{ZpCoupValue2} are very large due to the interference effects among them.

\begin{table}[t]
 \begin{center}
 \caption{\small Numerical results for the flavour-conserving $Z^{\prime}$ couplings $B_{uu}^{L,R}$ and $B_{dd}^{L,R}$ in the two different scenarios. The other captions are the same as in Table~\ref{ZpCoupValue1}.}
 \label{ZpCoupValue2}
 \vspace{0.2cm}
 \doublerulesep 0.8pt \tabcolsep 0.23in
 \begin{tabular}{lccccccccccc} \hline \hline
             &$B_{uu}^{L}$ &$B_{dd}^{L}$ &$B_{uu}^{R}$ &$B_{dd}^{R}$ \\\hline
  Scenario I   &$1.28\pm1.79$   &$1.85\pm1.87$   &$-0.20\pm0.54$  &$-0.20\pm0.53$ \\
  Scenario II &$0.55\pm0.25$   &$1.15\pm0.25$   &--- &---                 \\
 \hline \hline
 \end{tabular}
 \end{center}
\end{table}

\subsubsection*{Scenario~II: assuming that the right-handed flavour-conserving $Z^{\prime}$ couplings vanish}

\begin{figure}[t]
\begin{center}
\subfigure[]{\includegraphics [width=7.5cm]{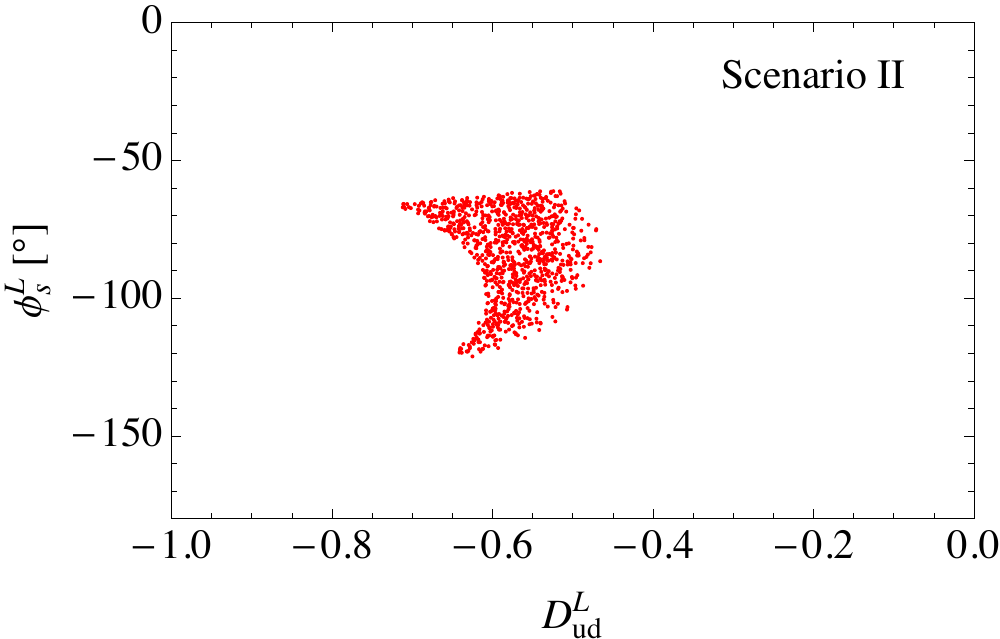}}\qquad
\subfigure[]{\includegraphics [width=7.5cm]{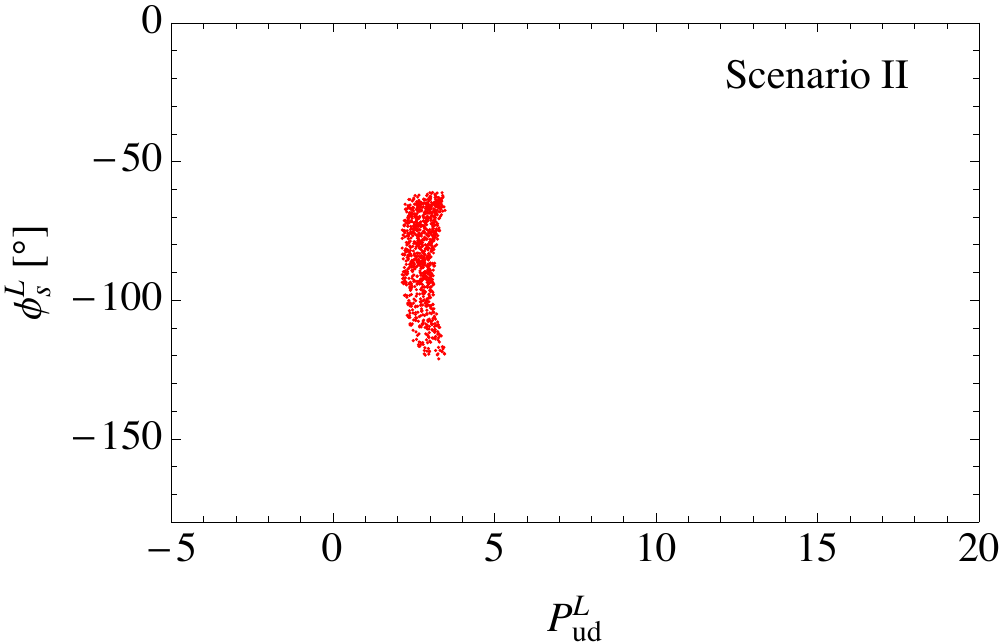}}
\caption{\label{ZpSpac2}\small The allowed regions for the $Z^{\prime}$ coupling parameters in scenario~II, under the constraints from the ${\cal B}(B\to \pi K,\rho K)$ and $A_{CP}(B\to \pi K,\rho K)$.}
\end{center}
\end{figure}

As has already been found in scenario~I, the left-handed flavour-conserving $Z^{\prime}$ couplings are crucial and non-negligible, whereas the right-handed ones are dispensable. Thus, as a maximally simplified case, in this scenario we assume that the right-handed $Z^{\prime}$ couplings vanish. Under the constraints from ${\cal B}(B\to \pi K,\rho K)$ and $A_{CP}(B\to \pi K,\rho K)$, the allowed parameter spaces of the flavour-conserving $Z^{\prime}$ couplings $D_{ud}^{L}$ and $P_{ud}^{L}$, as well as the weak phase $\phi^{L}_s$ are then shown in Fig.~\ref{ZpSpac2}. From the second row of Tables~\ref{ZpCoupValue1} and \ref{ZpCoupValue2}, in which the numerical results for $D_{ud}^{L}$, $P_{ud}^{L}$ and $B_{uu,dd}^{L}$ are presented respectively, one may find that, due to the absence of interference effects induced by the right ones, the uncertainties of the left-handed $Z^{\prime}$ couplings are significantly reduced. Moreover, it is found that the down-type coupling $B_{dd}^{L}\sim 1.2$ is about two times larger than the up-type one $B_{uu}^{L}\sim 0.6$.

Taking the fitted numerical results of the $Z^{\prime}$ couplings listed in Table~\ref{ZpCoupValue1} as inputs, we present our predictions for the observables in last two columns of Tables~\ref{pikbr}, \ref{pikdircp} and \ref{pikmixcp}. It can be seen that, compared to the SM prediction $\sim-11.7\%$, the direct CP asymmetry of $B^-\to\pi^0 K^-$ decay is significantly reduced by the $Z^{\prime}$ contribution, $(-0.5^{+4.2+1.0+3.3}_{-3.9-1.0-0.5})\%$ in scenario~I and $(-1.0^{+5.1+0.7+1.3}_{-3.7-0.8-1.4})\%$ in scenario~II, both of which are roughly consistent with the experimental data $(4.0\pm2.1)\%$, with their respective large uncertainties taken into account. In this sense, we say that the $Z^{\prime}$ model considered in this paper could provide a possible solution to the observed ``$\pi K$ CP puzzle". Meanwhile, it is also observed that our predictions for the other observables are in agreement with the experimental data within errors.

\subsection{$B_s\to K K$, $K K^{\ast}$ and $\pi^0 \phi$ decays}

Along with the successful running of LHCb, many decay modes of $B_s$ meson will be measured in the near future, which will provide another fertile ground to test the SM and various NP models. It is well-known that, under the approximation of neglecting annihilation contributions, some $B_s\to K K^{(\ast)}$ decays are related to the $B\to\pi K^{(\ast)}$ and $\rho K$ decays through the U-spin symmetry acting on the spectator quark of B meson. Especially the pairs ($\bar{B}_{s}\to{K}^{+}{K}^{-}$, $\bar{B}_{d}\to{\pi}^{+}{K}^{-}$) and ($\bar{B}_{s}\to{K}^{0}{\bar{K}}^{0}$, $B^-\to\pi^-{\bar K}^0$) are two interesting U-spin related examples. Moreover, these $B_s$ decay modes, being induced by the same quark-level $b\to s q\bar{q}$ transitions, involve the same $Z^{\prime}$ couplings as in $B_{u,d}\to\pi K^{(*)}$ and $\rho K$ decays. Thus, in this subsection, we shall investigate the impact of $Z^{\prime}$ contribution in $B_s\to K K$, $K K^{\ast}$ and $\pi^0 \phi$ decays.

\subsubsection{$B_s\to K K, K K^{\ast}$ decays}

The amplitudes of the two $B_s\to K K$ decays are given, respectively, as~\cite{Beneke3}
\begin{eqnarray}
{\cal A}_{\bar{B}_s^0\to\bar{K}^0 K^0}
   &=& B_{\bar{K} K} \Big[b_4^p - \frac{1}{2}b_{4,{\rm EW}}^p\Big]\nonumber\\
   &&+ A_{K \bar{K}}\Big[\alpha_4^p - \frac{1}{2}\alpha_{4,{\rm
    EW}}^p + \beta_3^p + \beta_4^p - \frac{1}{2}\beta_{3,{\rm EW}}^p - \frac{1}{2}\beta_{4,{\rm EW}}^p],
\label{amp5}
\end{eqnarray}
\begin{eqnarray}
{\cal A}_{\bar{B}_s^0\to K^- K^+}
   &=& B_{K^- K^+} \Big[
    \delta_{pu} b_1 + b_4^p + b_{4,{\rm EW}}^p\Big]\nonumber\\
   &&+ A_{ K^+ K^-}\Big[\delta_{pu} \alpha_1 + \alpha_4^{p}+ \alpha_{4,{\rm
    EW}}^p + \beta_3 + \beta_4 + \frac{1}{2}\beta_{3,{\rm EW}}^p - \frac{1}{2}\beta_{4,{\rm EW}}^p\Big]\,.
\label{amp6}
\end{eqnarray}
The corresponding amplitudes of $\bar{B}_s\to \bar{K} K^{\ast}$, $\bar{B}_s\to \bar{K}^{\ast} K$, $\bar{B}_s\to K^- K^{\ast+}$, and $\bar{B}_s\to K^{\ast+} K^-$ can be obtained from the above expressions with the replacement $(\bar{K} K)\to (\bar{K} K^{\ast})$, $(\bar{K} K)\to (\bar{K}^{\ast} K)$, $(K^- K^+)\to (K^- K^{\ast+})$, and $(K^- K^+)\to (K^{\ast-} K^+)$, respectively.

\begin{table}[t]
 \begin{center}
 \caption{\small The CP-averaged branching fractions~(in units of $10^{-6}$) of $\bar{B}_s \to K K^{(\ast)}$ and $\pi^0\phi$ decays. The other captions are the same as in Table~\ref{pikbr}.}
 \label{brKK}
 \vspace{0.2cm}
 \doublerulesep 0.8pt \tabcolsep 0.12in
 \begin{tabular}{lcccccccccccc} \hline \hline
 &\multicolumn{1}{c}{Decay} &\multicolumn{1}{c}{Exp.}& SM           &\multicolumn{2}{c}{$Z^{\prime}$ model} \\
  & Modes                                                      &data             &                              &Scenario I       &Scenario II           \\ \hline
 &$\bar{B}_s\to{K}^+K^-$                 &$24.5\pm1.8$ &$27.2^{+7.2+5.3}_{-6.0-3.8}$  &$26.9^{+7.8+5.3+1.6}_{-5.5-3.8-2.1}$ &$27.2^{+7.5+5.3+0.2}_{-5.9-3.8-0.2}$  \\
 &$\bar{B}_s\to{K}^0\bar{K}^0$      &$<66$           &$29^{+7+6}_{-6-4}$            &$30^{+7+6+1}_{-6-4-3}$             &$29^{+7+6+0}_{-6-4-0}$   \\
 \hline
 &$\bar{B}_{s}\to{K}^{+}K^{{\ast}-}$        &--- &$8.7^{+3.4+2.3}_{-2.4-1.6}$  &$9.0^{+3.1+2.3+0.4}_{-2.6-1.6-0.6}$ &$8.8^{+3.2+2.3+0.3}_{-2.4-1.6-0.2}$    \\
 &$\bar{B}_{s}\to{K}^{0}{\bar{K}}^{{\ast}0}$ &---  &$8.7^{+2.9+2.4}_{-2.0-1.7}$   &$8.8^{+2.6+4.0+0.6}_{-2.2-1.7-0.6}$  &$8.7^{+2.6+2.4+0.5}_{-1.9-1.6-0.5}$ \\
 &$\bar{B}_{s}\to{K}^{-}K^{{\ast}+}$                 &--- &$16.3^{+5.8+3.8}_{-4.7-2.7}$  &$18.1^{+6.3+4.0+1.8}_{-5.4-2.9-3.4}$ &$16.7^{+6.4+3.8+1.1}_{-4.6-2.7-1.1}$    \\
 &$\bar{B}_{s}\to{\bar{K}}^{0}{K}^{{\ast}0}$ &--- &$13.2^{+4.8+3.6}_{-3.5-2.5}$   &$13.8^{+4.7+3.7+1.7}_{-4.0-2.6-2.0}$  &$13.3^{+4.6+3.6+1.2}_{-3.5-2.5-1.3}$   \\\hline
  &$\bar{B}_{s}\to\pi^0\phi$        &---                  &$0.19^{+0.07+0.00}_{-0.05-0.00}$   &$0.38^{+0.10+0.01+0.36}_{-0.08-0.01-0.17}$  &$0.41^{+0.11+0.01+0.38}_{-0.09-0.01-0.22}$   \\
 \hline \hline
 \end{tabular}
 \end{center}
 \end{table}

\begin{table}[t]
 \begin{center}
 \caption{\small The direct CP asymmetries~(in unit of $10^{-2}$) of $\bar{B}_s \to K K^{(\ast)}$ and $\pi^0\phi$ decays. The other captions are the same as in Table~\ref{pikbr}.}
 \label{dcpKK}
 \vspace{0.2cm}
 \doublerulesep 0.8pt \tabcolsep 0.14in
 \begin{tabular}{lcccccccccccc} \hline \hline
 &\multicolumn{1}{c}{Decay}            &Exp.& SM                               &\multicolumn{2}{c}{$Z^{\prime}$ model} \\
       & Mode                                           &data            &                                  &Scenario I       &Scenario II           \\ \hline
 &$\bar{B}_{s}\to K^+K^-$                    &$2\pm18$ &$-15^{+3+1}_{-3-1}$     &$-16^{+3+0+3}_{-3-0-6}$ &$-15^{+3+0+0}_{-3-0-0}$   \\
 &$\bar{B}_{s}\to K^0\bar{K}^0$         &---          &$0.44^{+0.16+0.05}_{-0.13-0.05}$  &$0.45^{+0.17+0.05+0.07}_{-0.08-0.06-0.04}$ &$0.43^{+0.17+0.05+0.00}_{-0.14-0.05-0.00}$  \\
 \hline
 &$\bar{B}_{s}\to{K}^{+}K^{{\ast}-}$       &---  &$-45^{+9+5}_{-10-4}$  &$-60^{+10+4+16}_{-8-3-12}$ &$-56^{+10+4+1}_{-9-4-2}$   \\
 &$\bar{B}_{s}\to{K}^{0}{\bar{K}}^{{\ast}0}$ &---  &$0.5^{+0.2+0.1}_{-0.2-0.1}$   &$-13^{+2+1+21}_{-2-1-18}$  &$-5^{+2+1+1}_{-1-1-1}$     \\
 &$\bar{B}_{s}\to{K}^{-}K^{{\ast}+}$      &--- &$38^{+12+2}_{-10-3}$        &$37^{+10+1+1}_{-10-2-3}$ &$35^{+11+2+1}_{-11-2-1}$      \\
 &$\bar{B}_{s}\to{\bar{K}}^{0}{K}^{{\ast}0}$  &--- &$0.2^{+0.1+0.0}_{-0.1-0.0}$   &$4.4^{+2.8+0.9+11.8}_{-2.9-1.0-11.8}$  &$0.3^{+1.8+0.8+0.0}_{-2.2-0.9-0.0}$     \\\hline
  &$\bar{B}_{s}\to\pi^0\phi$        &---                  &$31^{+6+0}_{-6-0}$   &$8^{+5+0+6}_{-6-0-15}$  &$7^{+6+0+5}_{-6-0-11}$   \\
 \hline \hline
 \end{tabular}
 \end{center}
 \end{table}

\begin{table}[t]
 \begin{center}
 \caption{\small The mixing-induced CP asymmetries~(in unit of $10^{-2}$) of $\bar{B}_s \to K K$ and $\pi^0\phi$ decays. The other captions are the same as in Table~\ref{pikbr}.}
 \label{mcpKK}
 \vspace{0.2cm}
 \doublerulesep 0.8pt \tabcolsep 0.21in
\begin{tabular}{lccccccc} \hline \hline
 &\multicolumn{1}{c}{Decay}    &\multicolumn{1}{c}{Exp.}    &SM           &\multicolumn{2}{c}{$Z^{\prime}$ model}\\
                                    &Mode     &data                     &                                   &Scenario I  &Scenario II \\ \hline
 &$\bar{B}_{s}\to{K}^{+}{K}^{-}$     &$17\pm19$       &$25^{+6+2}_{-6-2}$ &$13^{+7+2+25}_{-7-2-23}$ &$23^{+6+2+1}_{-6-2-1}$   \\
 &$\bar{B}_{s}\to{K}^{0}{\bar{K}}^{0}$   &---           &$0.6^{+0.2+0.0}_{-0.1-0.0}$  &$0.6^{+0.1+0.0+0.0}_{-0.1-0.0-0.1}$ &$0.6^{+0.2+0.0}_{-0.1-0.0}$ \\\hline
  &$\bar{B}_{s}\to\pi^0\phi$        &---                  &$39^{+10+4}_{-11-3}$   &$-99^{+1+0+68}_{-0-0-1}$  &$-99^{+2+0+46}_{-0-0-1}$   \\
 \hline \hline
 \end{tabular}
 \end{center}
 \end{table}

With the theoretical inputs listed in the Appendix and the restricted parameter spaces of the $Z^{\prime}$ couplings listed in Table~\ref{ZpCoupValue1}, our predictions for the branching fractions, the direct and the mixing-induced CP asymmetries of $B_s\to K K$, $K K^{\ast}$ and $\pi^0 \phi$ decays are summarized, respectively, in Tables~\ref{brKK}, \ref{dcpKK} and \ref{mcpKK}. Among these decays, only the $\bar{B}_{s}\to{K}^{+}{K}^{-}$ decay has been measured so far, for which our SM prediction is consistent with the previous theoretical evaluations~\cite{Cheng2,VirtoKK} and agrees well with the experimental data.

For the penguin-dominated $B_s\to K K$ and $K K^{\ast}$ decays, as the branching fractions are dominated by the module of the effective coefficients $\alpha_{4}$ and $\alpha_{4,{\rm EW}}$, to which the $Z^{\prime}$ contributions are colour-suppressed, the NP effect in the branching fractions are not significant and diluted by the large theoretical uncertainties. On the other hand, due to the weak phase $\phi_s^L$ being non-zero, some of their CP asymmetries are very sensitive to the $Z^{\prime}$ contributions, which can be clearly seen from Tables~\ref{dcpKK} and \ref{mcpKK}.

\begin{figure}[t]
\begin{center}
\subfigure[]{\includegraphics[width=7.0cm]{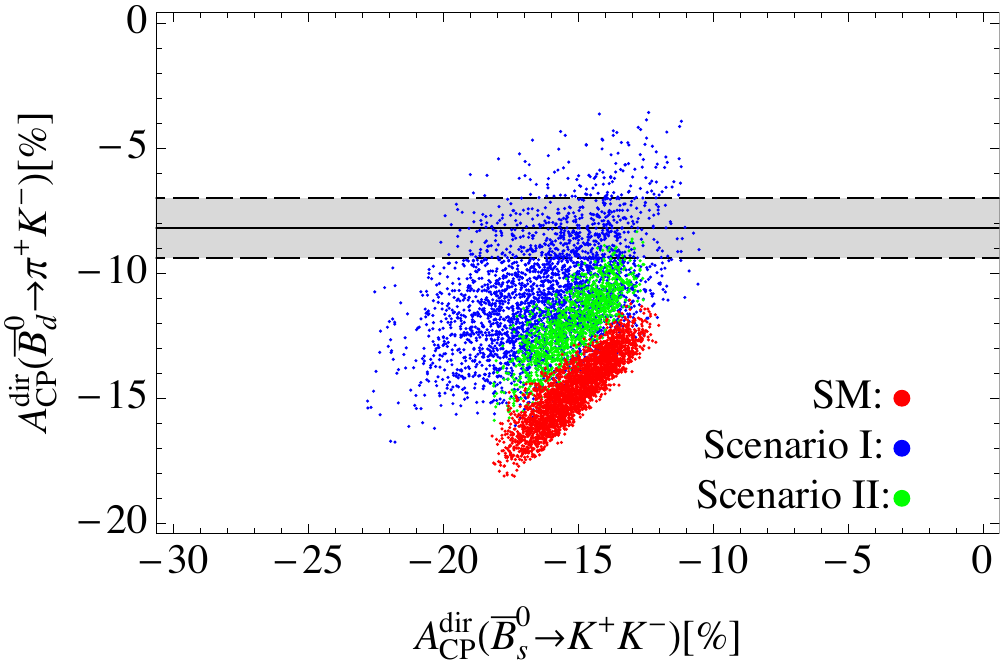}}\qquad
\subfigure[]{\includegraphics[width=7.0cm]{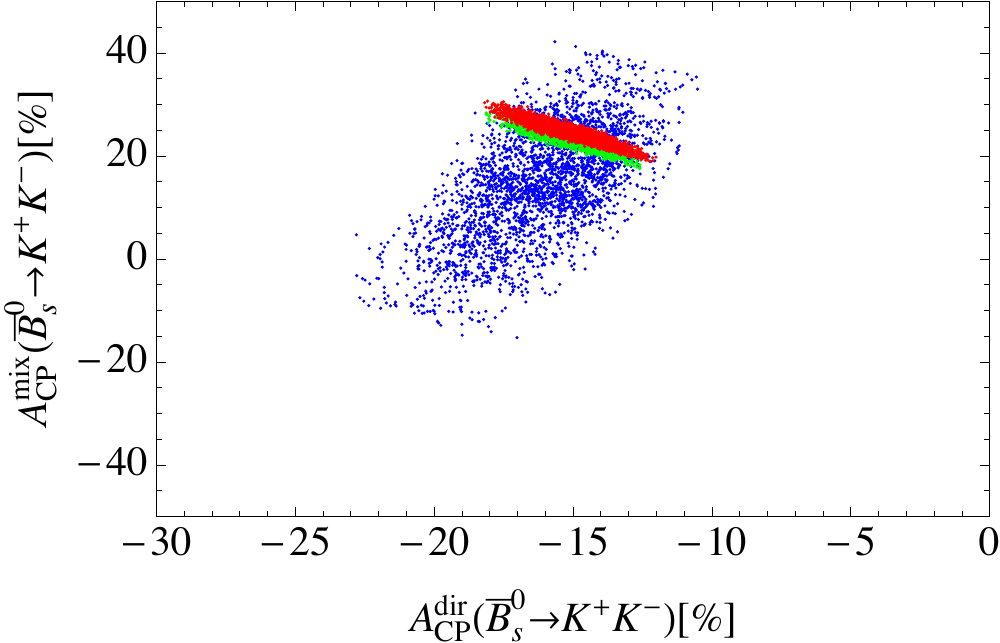}}\\
\subfigure[]{\includegraphics[width=7.0cm]{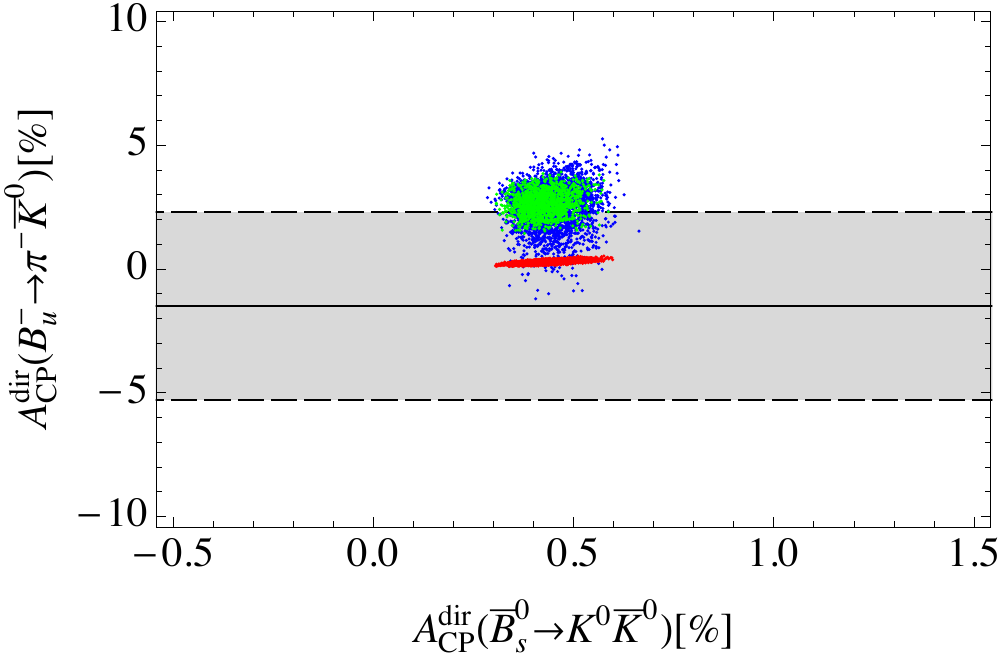}}\qquad
\subfigure[]{\includegraphics[width=7.0cm]{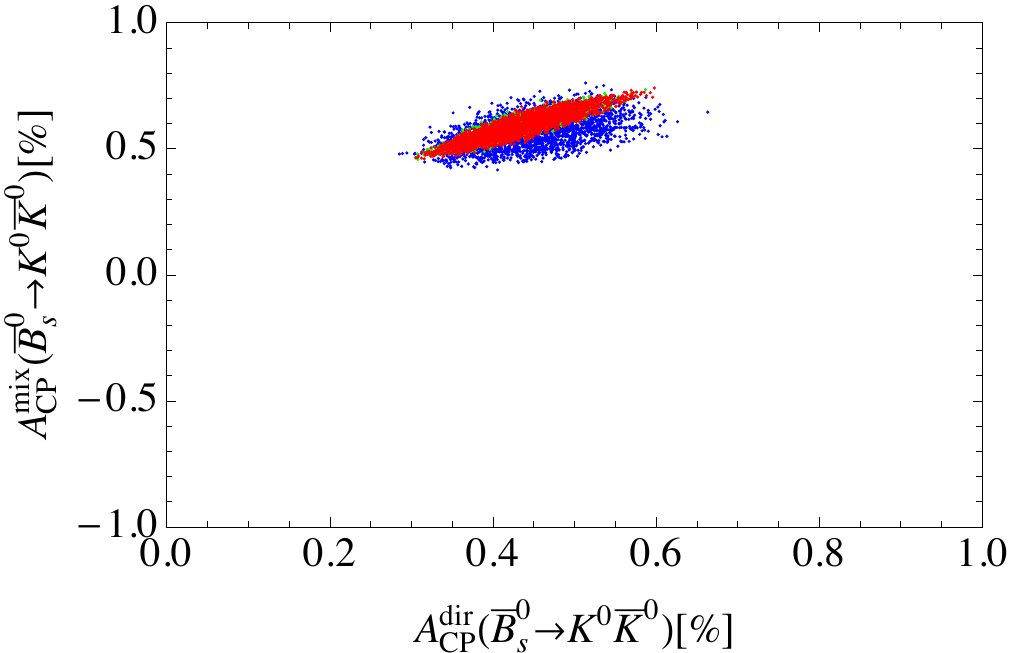}}
\caption{\label{kkvs} \small Correlations between (a) $A_{CP}^{dir}(\bar{B}_{s}\to K^+K^-)$ versus $A_{CP}^{dir}(\bar{B}_{d}\to \pi^+K^-)$,~(b) $A_{CP}^{dir}(\bar{B}_{s}\to K^+K^-)$ versus $A_{CP}^{mix}(\bar{B}_{s}\to K^+K^-)$,~(c) $A_{CP}^{dir}( \bar{B}_s^0\to\bar{K}^0 K^0)$ versus $A_{CP}^{dir}( B^-\to\pi^- K^0)$, and~(d) $A_{CP}^{dir}( \bar{B}_s^0\to\bar{K}^0 K^0)$ versus $A_{CP}^{mix}( \bar{B}_s^0\to\bar{K}^0 K^0)$, both within the SM and in the $Z^{\prime}$ model with the two different scenarios. The gray bands denote the experimental data within $2\sigma$ error bars.}
\end{center}
\end{figure}

In order to further test the $Z^{\prime}$ effects and check if the $Z^{\prime}$ contributions in the two different scenarios could be distinguished from each other and from the SM predictions, we show in Fig.~\ref{kkvs} the correlations between various CP asymmetries both within the SM and in the $Z^{\prime}$ model, in which the theoretical uncertainties induced by the input parameters listed in Appendix are considered. It is observed that, for $A_{CP}^{dir}(\bar{B}_{d}\to \pi^+K^-)$ and $A_{CP}^{dir}(\bar{B}_{s}\to K^+K^-)$, while the results in the case of scenario~II are quite similar to the SM ones, the scenario~I case could deviate significantly from the SM predictions, which could provide a useful probe of $Z^{\prime}$ contribution with right-handed $u(d)-u(d)-Z^{\prime}$ couplings. For $A_{CP}^{mix}(\bar{B}_{s}\to K^+K^-)$, the $Z^{\prime}$ effect is even more significant and could flip the sign of the SM prediction. However, the experimental data for these asymmetries are currently still too rough to give a definite conclusion. In addition, even though the $Z^{\prime}$ contributions also exhibit some deviations from the SM predictions, the observables $A_{CP}^{dir, mix}( \bar{B}_s^0\to\bar{K}^0 K^0)$ are too small to be accessible  in the near future.

\begin{figure}[t]
\begin{center}
\subfigure[]{\includegraphics[width=7.4cm]{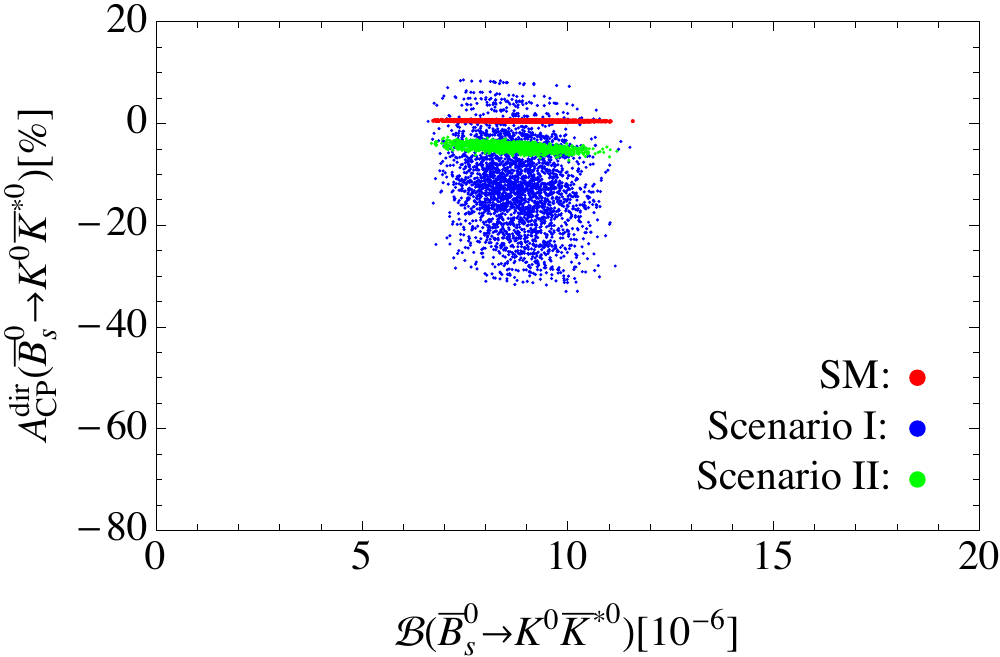}}\qquad
\subfigure[]{\includegraphics[width=7.5cm]{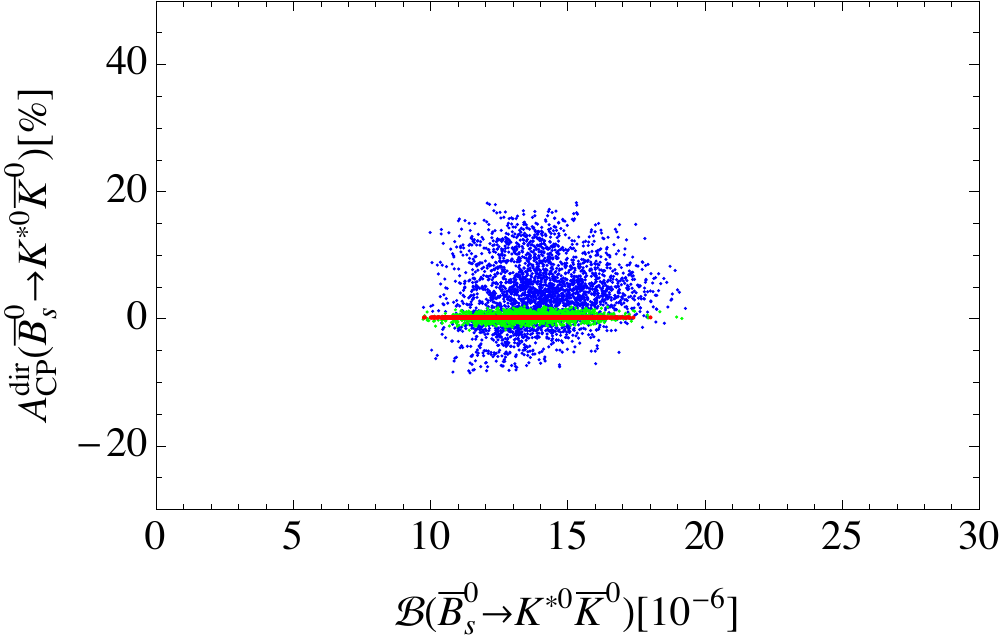}}
\caption{\label{kkvvs}\small (a) ${\cal B}(\bar{B}_s\to{K}^{0}\bar{K}^{\ast 0})$ versus $A_{CP}^{dir}(\bar{B}_s\to{K}^{0}\bar{K}^{\ast 0})$,  (b) ${\cal B}(\bar{B}_s\to{K}^{\ast 0}\bar{K}^{ 0})$ versus $A_{CP}^{dir}(\bar{B}_s\to{K}^{\ast 0}\bar{K}^{0})$ both within the SM and in the $Z^{\prime}$ model with the two different scenarios.}
\end{center}
\end{figure}

It is also found that $A_{CP}^{dir}(\bar{B}_s\to{K}^{0}\bar{K}^{\ast 0})$ and $A_{CP}^{dir}(\bar{B}_s\to{K}^{\ast 0}\bar{K}^{0})$ are another two interesting observables that can be used to probe the $Z^{\prime}$ effect. As is shown in Fig.~\ref{kkvvs}, while both of them are predicted to be around zero within the SM, the $Z^{\prime}$ contribution in scenario~I could bring a significant deviation from the SM prediction. Explicitly, in this scenario, both a large negative $A_{CP}^{dir}(\bar{B}_s\to{K}^{0}\bar{K}^{\ast 0})$ and a large positive $A_{CP}^{dir}(\bar{B}_s\to{K}^{\ast 0}\bar{K}^{0})$ are predicted, which, if confirmed by future experimental measurements, could be used to distinguish these two different scenarios from each other.

\subsubsection{$\bar{B}_{s}\to\pi^0\phi$ decay}

Besides $B_s\to K K$ and $K K^{\ast}$ decays, the $\bar{B}_s \to \pi^0 \phi$ decay is another interesting and important process to probe the $Z^{\prime}$ effect~\cite{ZpLiphipi}, even though being very rare with a branching fraction of order of $10^{-7}$. The decay amplitude of this mode is very simple and given as~\cite{Beneke3}
\be
{\cal A}_{\bar{B}_s\to \pi^0 \phi}=\frac{A_{\phi \pi}}{\sqrt{2}}\Big[\delta_{pu} \alpha_2 +\frac{3}{2}\alpha_{3,{\rm EW}}^p\Big]\,.
\label{amp7}
\ee

From Eq.~(\ref{amp7}), it can be seen that this decay is dominated by the effective EW-penguin coefficient $\alpha_{3,{\rm EW}}^p=a_9^p-a_7^p$, while the contribution from $\alpha_2$ is CKM-suppressed. Recalling that the significant $Z^{\prime}$ effect on $A_{CP}^{dir}(B^-\to\pi^0K^-)$ is through $\alpha_{3,{\rm EW}}^p$~(see Eq.~(\ref{c79})), it is therefore expected that the observables of $\bar{B}_s\to \pi^0 \phi$ decay should also be very sensitive to the $Z^{\prime}$ contribution. Another important feature of this decay channel is that it is not contaminated by the annihilation correction, which suffers from large theoretical uncertainties due to the known end-point divergence. Thus, the $\bar{B}_s \to \pi^0 \phi$ decay is considered as a relatively ``clean" channel for testing the SM and probing possible NP effects~\cite{Vernazza}.

\begin{figure}[t]
\begin{center}
\subfigure[]{\includegraphics[width=7.5cm]{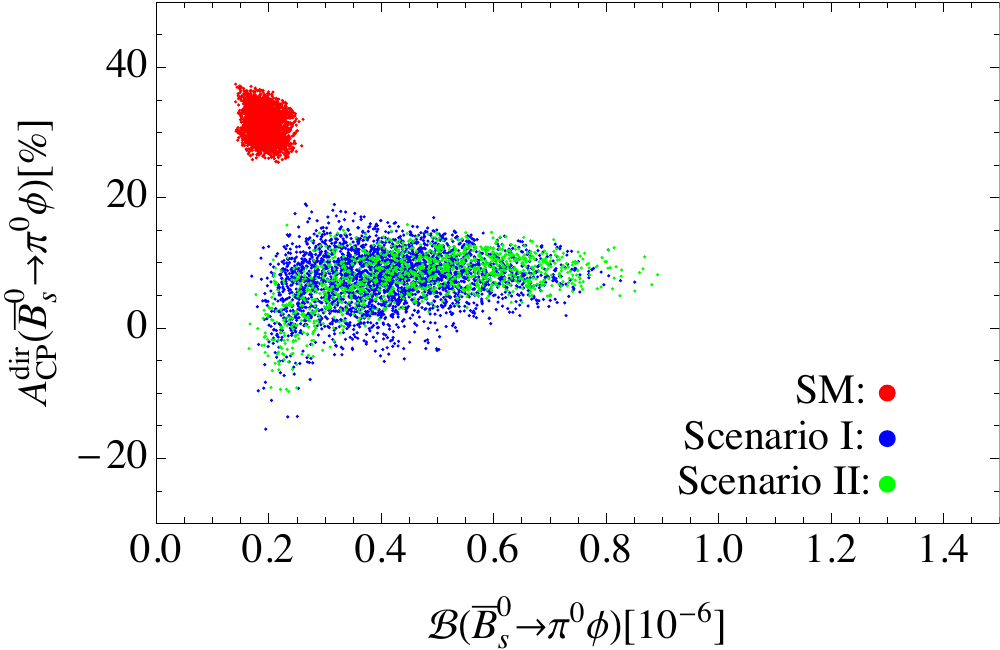}}\qquad
\subfigure[]{\includegraphics[width=7.5cm]{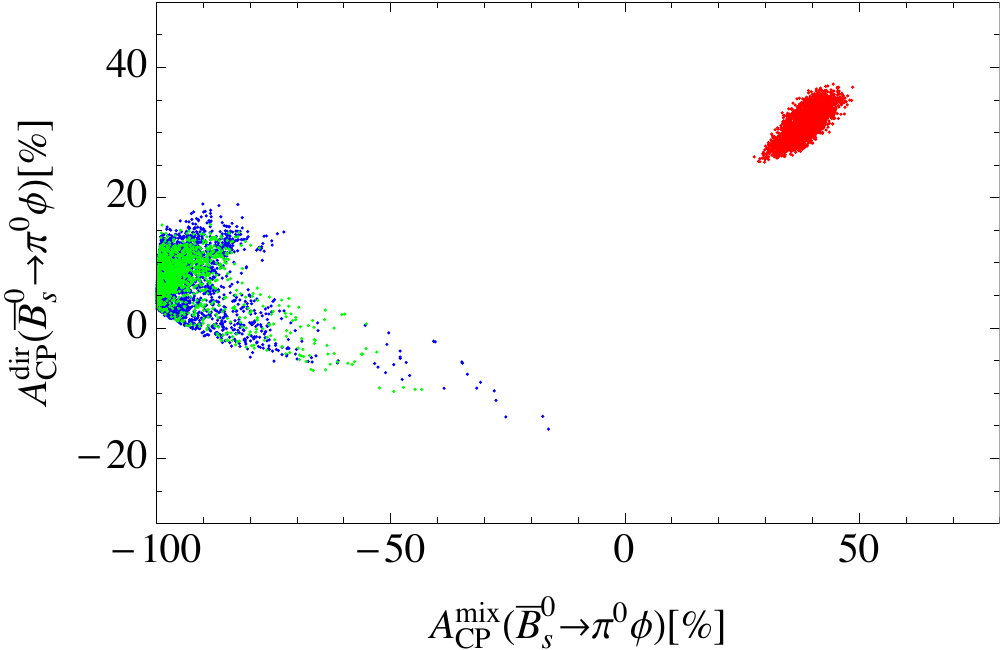}}
\caption{\label{phipivs}\small ~(a) ${\cal B}(\bar{B}_s\to \pi^0 \phi)$ versus $A_{CP}^{dir}(\bar{B}_s\to \pi^0 \phi)$, and~(b) $A_{CP}^{mix}(\bar{B}_s\to \pi^0 \phi)$ versus $A_{CP}^{dir}(\bar{B}_s\to \pi^0 \phi)$ both within the SM and in the $Z^{\prime}$ model with scenario~I and scenario~II.}
\end{center}
\end{figure}

With the obtained $Z^{\prime}$ coupling parameters listed in Table~\ref{ZpCoupValue1}, our predictions for the observables of $\bar{B}_s \to \pi^0 \phi$ decay both within the SM and in the $Z^{\prime}$ model with the two different scenarios are presented in the last row of Tables~\ref{brKK}, \ref{dcpKK} and \ref{mcpKK}. As has already been mentioned, the $Z^{\prime}$ corrections are significant for all of these observables compared to the SM predictions. Numerically, one can find that ${\cal B}(\bar{B}_s\to \pi^0 \phi)$ could be enhanced to $\sim 0.8\times 10^{-6}$ by the $Z^{\prime}$ contribution, which is about four times larger than the SM prediction $\sim 0.19\times 10^{-6}$. For $A_{CP}^{dir}(\bar{B}_s\to \pi^0 \phi)$, the difference between the SM and the $Z^{\prime}$ predictions $A_{CP}^{dir,SM}-A_{CP}^{dir,Z^{\prime}}$ is about $(10\sim30)\%$. For $A_{CP}^{mix}(\bar{B}_s\to \pi^0 \phi)$, the difference is even larger, with signs completely flipped. Moreover, as is shown in Fig.~\ref{phipivs}, even with the theoretical uncertainties taken into account, the predicted $A_{CP}^{dir,mix}(\bar{B}_s\to \pi^0 \phi)$ in the $Z^{\prime}$ model deviates entirely from the SM regimes, which means that such an observable is very powerful for probing possible $Z^{\prime}$ effects. However, the two different $Z^{\prime}$ scenarios are found to be almost indistinguishable from each other by this decay. Thus, future experimental measurements of $\bar{B}_s\to \pi^0 \phi$ decay, especially the direct and mixing-induced CP asymmetries, will play a very important role in confirming or refuting the possible $Z^{\prime}$ effect considered in this paper.

As is discussed extensively in the literature~\cite{Barger,changpikzp,pikpuz}, in order to reconcile the observed ``$\pi K$ CP puzzle", one has to modify either the color-suppressed tree amplitude $\alpha_2^p$ or the EW-penguin amplitude $\alpha_{3,{\rm EW}}^p$; the way studied in this paper by introducing a family non-universal $Z^{\prime}$ model belongs to the latter. It is interesting to note that the decay amplitude ${\cal A}_{\bar{B}_s\to \pi^0 \phi}$~(see Eq.~(\ref{amp7})) involves both of these two effective coefficients. Thus, with the coming experimental measurements of $\bar{B}_s \to \pi^0 \phi$ decay at LHCb and Super-KEKB, a combined study of $B\to\pi K$ and $\bar{B}_s\to \pi^0 \phi$ decays will provide a much more crucial test of various NP scenarios designed to resolve the observed ``$\pi K$ CP puzzle".

\subsection{$B\to \phi K$ decays}

In a family non-universal $Z^{\prime}$ model, the couplings of $Z^{\prime}$ boson to quarks are generally not the same for different generations. Focusing on the hadronic B-meson decays induced by the quark-level $b\to s$ transitions, this means that the flavour-conserving $s-s-Z^{\prime}$ coupling might be different from the $d-d-Z^{\prime}$ one discussed in the previous two subsections. In order to further test such a $Z^{\prime}$ model, in this subsection, we shall proceed to discuss the penguin-dominated $B\to \phi K$ decays, which are induced by the quark-level $b\to s s\bar{s}$ transition and hence offer access to the strength of $s-s-Z^{\prime}$ coupling.

With the input parameters summarized in the Appendix, our SM predictions for the branching fractions, direct and mixing-induced CP asymmetries of $B\to\phi K$ decays are presented in the third column of Table~\ref{phiKobs}. It can be seen that, while most of these observables are consistent with the experimental data, our estimation of $A_{CP}^{dir}(B^-\to\phi K^-)=(0.4^{+0.2}_{-0.2})\%$, although being in good agreement with the previous SM predictions~(for instance, $0.7\%$~(QCDF, S4)~\cite{Beneke3} and $(1^{+0}_{-1})\%$~(pQCD)~\cite{LiPV}), is still about $2.4\sigma$ smaller than the experimental data~\cite{HFAG}
\be\label{phiKHFAG}
A_{CP}^{dir}(B^-\to \phi K^-)=(10.4\pm4.2)\%\,,
\ee
which is obtained by taking average over the following experimental data
\ba
A_{CP}^{dir}(B^-\to \phi K^-)=
\left\{\begin{array}{l}
(-7\pm17^{+3}_{-2})\%\qquad{\rm CDF~\cite{phiKCDF}}\\
(1\pm12\pm5)\%\qquad{\rm Belle~\cite{phiKBelle}}\\
(12.8\pm4.4\pm1.3)\%\qquad{\rm BaBar\,~\cite{phiKBABAR}}\,,
\end{array}\right.
\ea
and is obviously dominated by the BaBar measurement. Recently, using the known value of the $B^-\to J/\psi K^-$ asymmetry, $A_{CP}^{dir}(B^-\to \phi K^-)$ has also been measured by the LHCb collaboration and is determined to be $(2.2\pm2.1\pm0.9)\%$~\cite{phiKLHCb}, which is in agreement with the SM prediction but is not included in the HFAG's average. Averaging the BaBar and the LHCb data roughly, we get the weighted average $A_{CP}^{dir}(B^-\to \phi K^-)=(4.3\pm 2.0)\%$, which is still about $2.2\sigma$ away from the SM expectation. So, such a possible discrepancy, if confirmed by more precise experimental measurements, would imply possible new sources of CP violation beyond the SM. In the following, we shall investigate whether the family non-universal $Z^{\prime}$ model considered in this paper could provide a possible solution.

\begin{table}[t]
 \begin{center}
 \caption{\small The CP-averaged branching ratios~(in units of $10^{-6}$), the direct and the mixing-induced CP asymmetries~(in units of $10^{-2}$) of $B\to \phi K$ decays both within the SM and in the $Z^{\prime}$ model with the two different scenarios. The other captions are the same as in Table~\ref{pikbr}.}
 \label{phiKobs}
 \vspace{0.2cm}
 \doublerulesep 0.8pt \tabcolsep 0.16in
 \begin{tabular}{lcccccccccccc} \hline \hline
 \multicolumn{1}{c}{Observables}       &\multicolumn{1}{c}{Exp.}&\multicolumn{1}{c}{ SM }&\multicolumn{2}{c}{$Z^{\prime}$ model} \\
                                                                 &     data                                 &                                &Scenario I             &Scenario II      \\ \hline
 ${\cal B}(B^-\to\phi K^-)$                 &$8.8\pm0.5$                &$10.0^{+3.4+2.7}_{-2.8-1.8}$  &$10.0^{+3.1+2.6+0.4}_{-2.7-1.8-0.7}$ &$10.0^{+3.1+2.7+0.7}_{-2.7-1.8-0.6}$\\
 ${\cal B}({\bar B}^0\to\phi {\bar K}^0)$ &$7.3^{+0.7}_{-0.6}$&$9.1^{+3.2+2.4}_{-2.6-1.7}$   &$9.2^{+2.9+2.4+0.5}_{-2.5-1.7-1.8}$  &$9.1^{+2.9+2.5+0.7}_{-2.5-1.6-0.6}$  \\\hline
 $A_{CP}^{dir}(B^-\to\phi K^-)$                      &$10.4\pm4.2$                &$0.5^{+0.3+0.0}_{-0.2-0.0}$  &$12.7^{+1.7+0.7+6.1}_{-1.7-0.8-11.4}$ &$11.3^{+1.0+0.7+7.5}_{-1.0-0.9-9.2}$\\
 $A_{CP}^{dir}({\bar B}^0\to\phi {\bar K}^0)$ &$-1\pm14$  &$0.8^{+0.4+0.1}_{-0.3-0.1}$   &$8^{+1+1+14}_{-1-1-19}$            &$12^{+1+1+8}_{-1-1-9}$  \\\hline
 $A_{CP}^{mix}({\bar B}^0\to\phi {\bar K}^0)$ &$74^{+11}_{-13}$  &$82^{+8+0}_{-17-0}$   &$91^{+5+1+7}_{-7-1-11}$     &$86^{+6+1+2}_{-9-1-4}$  \\
 \hline \hline
 \end{tabular}
 \end{center}
 \end{table}

\begin{figure}[t]
\begin{center}
\subfigure[]{\includegraphics[width=7.0cm]{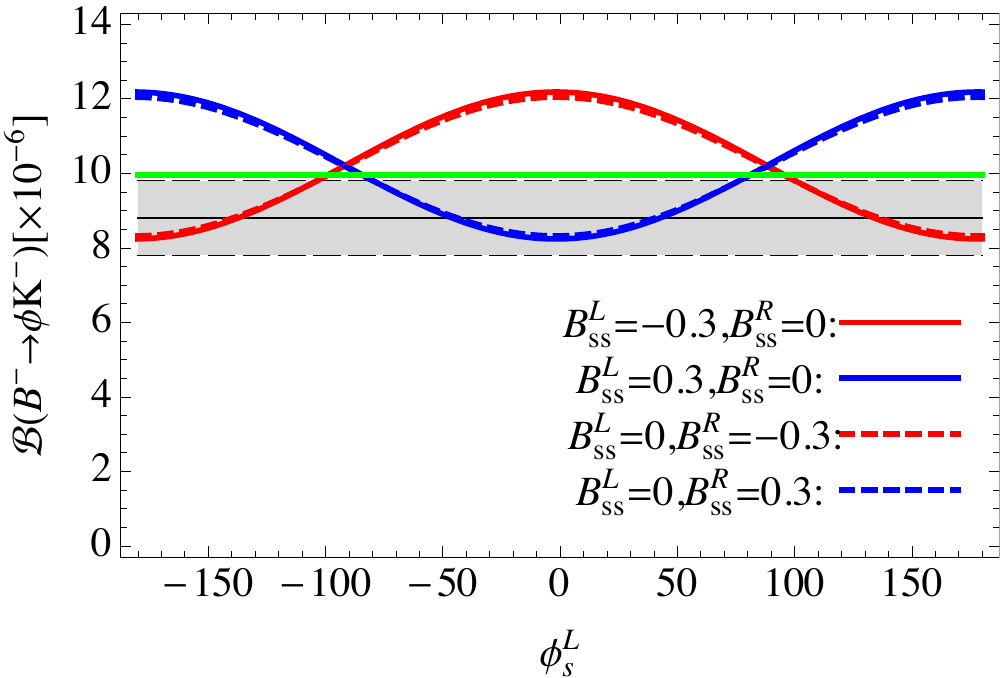}}\qquad
\subfigure[]{\includegraphics[width=7.3cm]{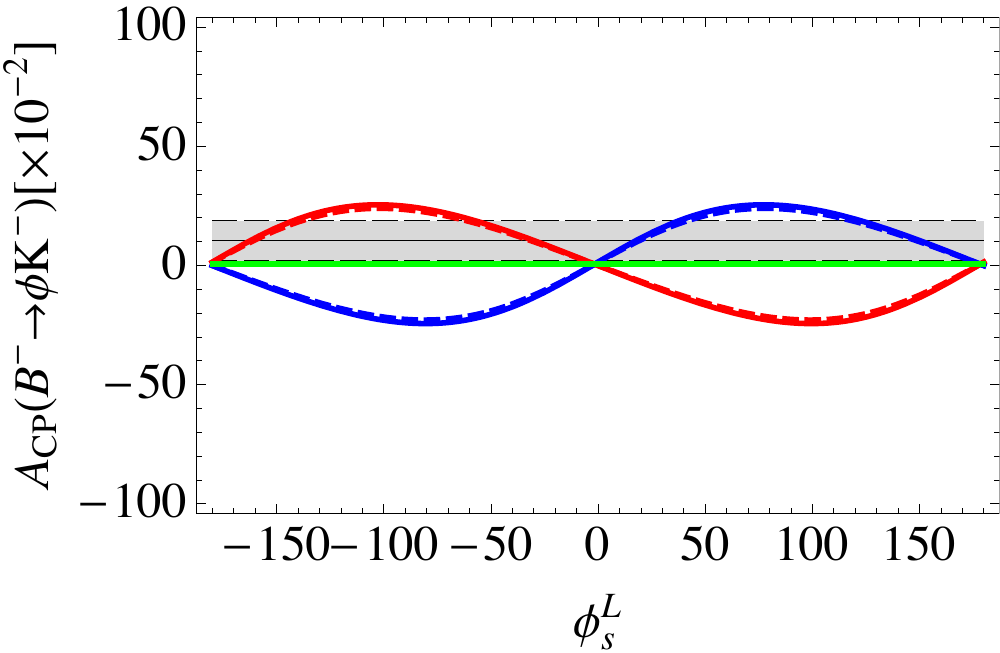}}\\
\subfigure[]{\includegraphics[width=7.0cm]{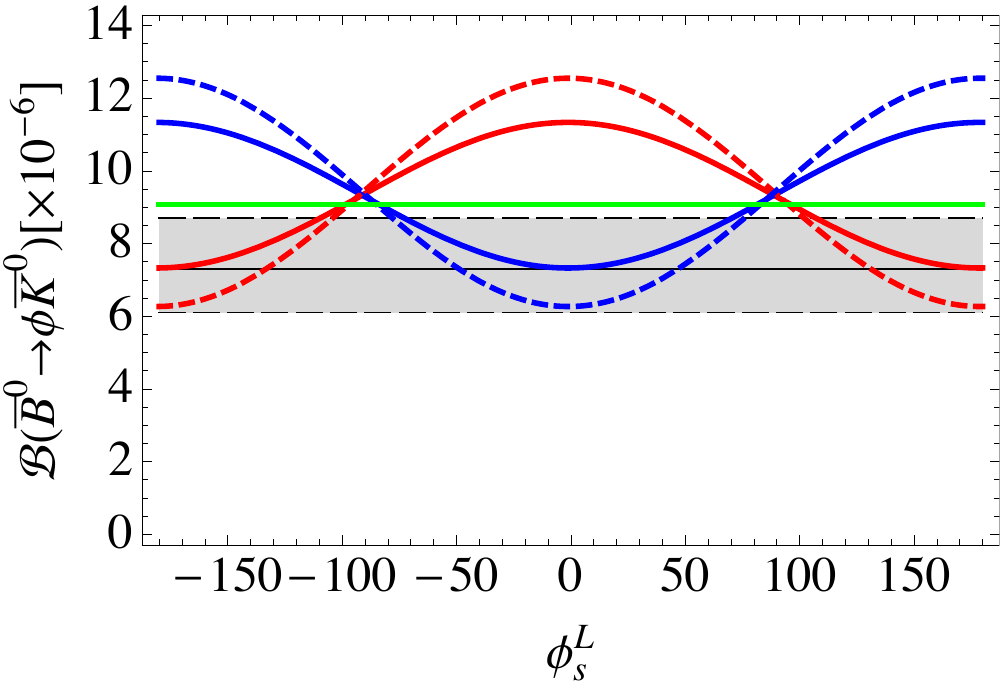}}\qquad
\subfigure[]{\includegraphics[width=7.3cm]{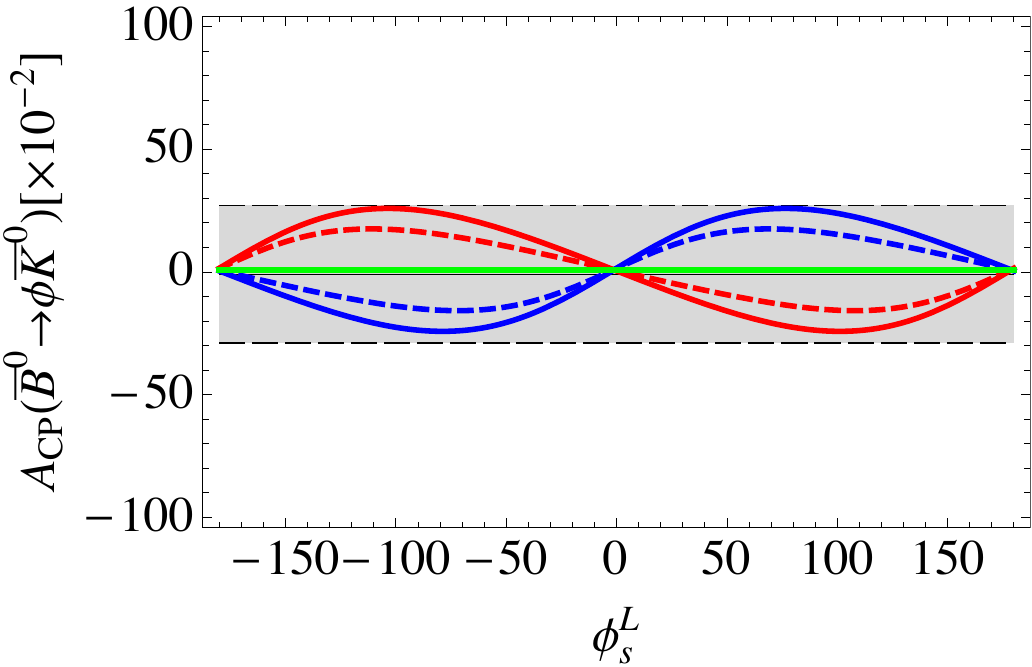}}
\caption{\label{phiK_zp}\small The dependence of observables in $B{\to}\phi K$ decays on the weak phase $\phi_s^L$, with $|B_{sb}^L|=0.5\times10^{-3}$ and $B_{ss}^{L,R}$ values labeled in Fig.~\ref{phiK_zp}(a). The experimental data and the SM predictions are shown as shaded regions~(within $2\sigma$ error-bars) and green lines, respectively.}
\end{center}
\end{figure}

The dependence of the observables of $B{\to}\phi K$ decays on the $Z^{\prime}$ coupling parameters are shown in Fig.~\ref{phiK_zp}. It is found that, with $\phi_s^L\sim-91^{\circ}$ fitted from $B\to\pi K$ decays, the $Z^{\prime}$ contributions with a negative $B_{ss}^L$ and/or $B_{ss}^R$ are helpful to moderate the discrepancy for $A_{CP}^{dir}(B^-\to\phi K^-)$, which is shown in Fig.~\ref{phiK_zp}(b). At the same time, such a possible solution also satisfies the constraint from $A_{CP}^{dir}({\bar B}^0\to\phi {\bar K}^0)$, as is shown in Fig.~\ref{phiK_zp}(d). However, from Figs.~\ref{phiK_zp}(a) and \ref{phiK_zp}(c), one can find that this solution is marginal around $\phi_s^L\sim-91^{\circ}$. An exact numerical evaluation is, therefore, needed to find the allowed regions for the $Z^{\prime}$ coupling parameters, which will be presented in the following.

\subsubsection*{Scenario II: assuming that the right-handed flavour-conserving $Z^{\prime}$ couplings vanish}

\begin{figure}[t]
\begin{center}
\includegraphics[width=7.5cm]{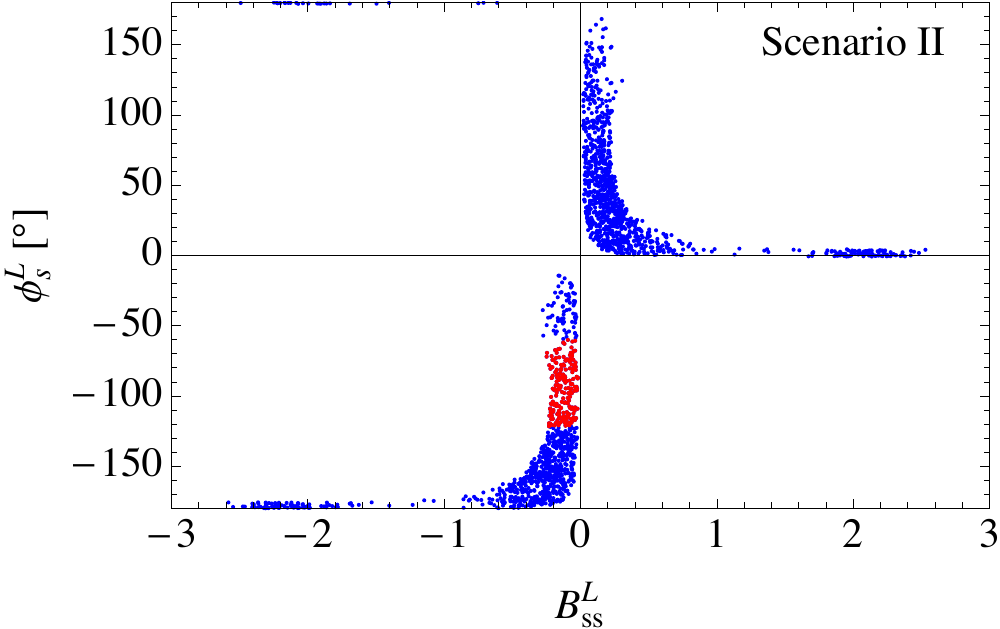}
\caption{\label{ZpSpac3}\small The allowed regions~(blue) for the $Z^{\prime}$ coupling parameters $B_{ss}^L$ and $\phi_s^L$ under the $2\sigma$ constraints of the branching ratios and direct CP asymmetries of $B\to\phi K$ decays. The red region is obtained with $\phi_s^L=-91^{\circ}\pm31^{\circ}$ fitted from $B\to \pi K$, $\pi K^{\ast}$ and $\rho K$ decays.}
\end{center}
\end{figure}

Firstly, we study the case with only left-handed $s-s-Z^{\prime}$ coupling being nonzero. Under the constraints from $B^-\to\phi K^-$ and ${\bar B}^0\to\phi {\bar K}^0$ decays, the allowed regions~(blue) are shown in Fig.~\ref{ZpSpac3}. It is found that there exist two separated allowed regions, $B_{ss}^L<0$ with $\phi_s^L\in[-\pi,0]$ and $B_{ss}^L>0$ with $\phi_s^L\in[0,\pi]$, mainly due to the constraint from $A_{CP}^{dir}(B^-\to\phi K^-)$ as has already been shown in Fig.~\ref{phiK_zp}(b). However, taking into account the bound on the weak phase $\phi_s^L$ from $B\to \pi K$, $\pi K^{\ast}$ and $\rho K$ decays, $\phi_s^L=-91^{\circ}\pm31^{\circ}$, the allowed regions will be significantly reduced and are shown in red in Fig.~\ref{ZpSpac3}. The corresponding numerical results are given in Table~\ref{ZpPareValueSS}, in which a negative $B_{ss}^{L}=-0.13\pm0.11$ with $\phi_s^L=-91^{\circ}\pm31^{\circ}$ is needed to moderate the large divergency for $A_{CP}^{dir}(B^-\to\phi K^-)$. Comparing this result with the value of $B_{dd}^{L}$ listed in table~\ref{ZpCoupValue2}, one can easily find that $B_{ss}^{L}\neq B_{dd}^{L}$ in this scenario.

\begin{table}[t]
 \begin{center}
 \caption{\small Numerical results for the flavour-conserving $Z^{\prime}$ parameters $B_{ss}^L$ and $B_{ss}^R$ in two different scenarios, with $\phi_s^L$ fitted from $B\to \pi K$, $\pi K^{\ast}$ and $\rho K$ decays.}
 \label{ZpPareValueSS}
 \vspace{0.2cm}
 \doublerulesep 0.8pt \tabcolsep 0.4in
 \begin{tabular}{lccccccccccc} \hline \hline
             &$B_{ss}^{L}$ &$B_{ss}^{R}$ &$\phi^{L}_s[^{\circ}]$\\\hline
Scenario I   &$0.05\pm0.54$   &$-0.21\pm0.51$  &$-91\pm33$\\
Scenario II  &$-0.13\pm0.11$   &---  &$-91\pm31$\\
 \hline \hline
 \end{tabular}
 \end{center}
 \end{table}

With the obtained numerical results listed in Table~\ref{ZpPareValueSS} as inputs, we present our predictions in the fifth column of Table~\ref{phiKobs}. One can find that our theoretical prediction $A_{CP}^{dir}(B^-\to\phi K^-)=(11.3^{+1.0+0.7+7.5}_{-1.0-0.9-9.2})\%$ is in agreement with the experimental data $A_{CP}^{dir}(B^-\to\phi K^-)=(10.4\pm4.2)\%$~\cite{HFAG} at $1\sigma$ level. In addition, our predictions for the other observables also agree well with the current experimental measurements.

\subsubsection*{Scenario~I: without any simplifications for the flavour-conserving $Z^{\prime}$ couplings}

\begin{figure}[t]
\begin{center}
\subfigure[]{\includegraphics [width=7.5cm]{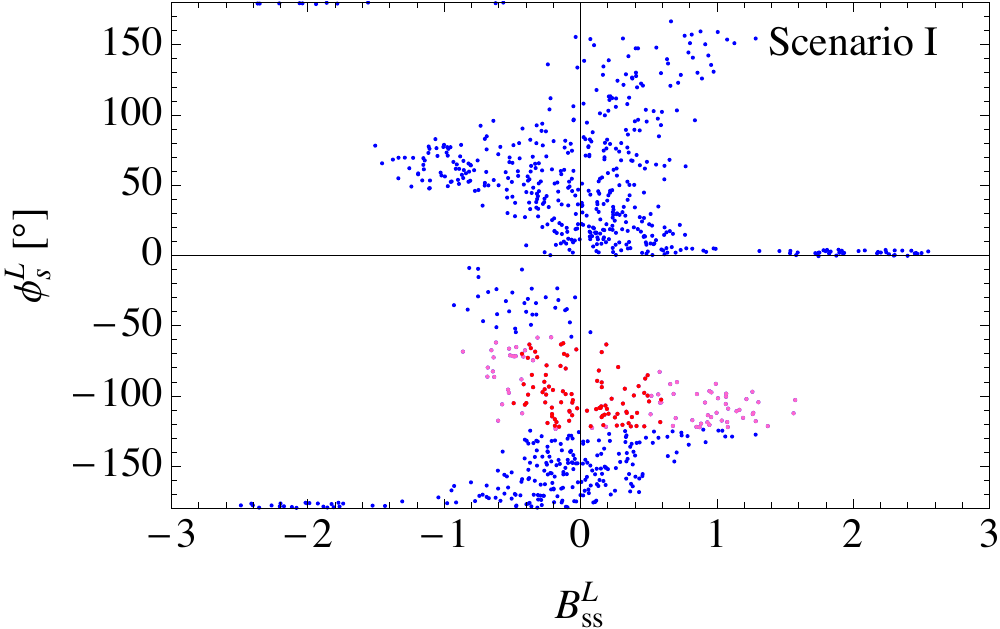}}\qquad
\subfigure[]{\includegraphics [width=7.5cm]{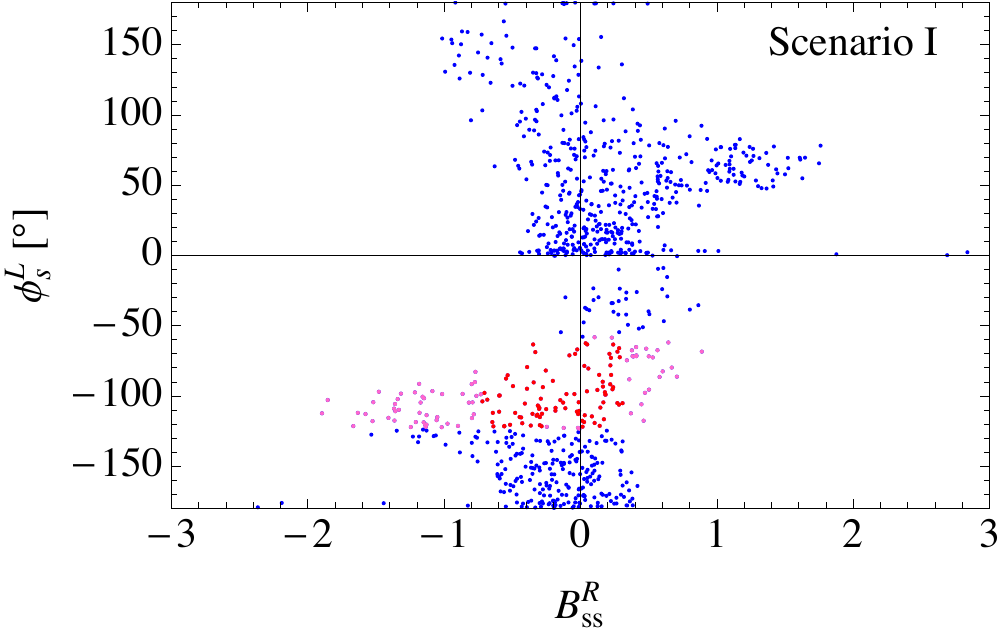}}
\caption{\label{ZpSpac4}\small Allowed regions for the parameters $B_{ss}^L$, $B_{ss}^R$ and $\phi_s^L$ under the constraints of the branching ratios and direct CP asymmetries of $B\to\phi K$ decays. See text for the colour interpretations.}
\end{center}
\end{figure}

From Fig.~\ref{phiK_zp}(b), it can be seen that a negative $B_{ss}^R$ with $\phi_s^L\sim-91^{\circ}$ is also preferred to reconcile the discrepancy of $A_{CP}^{dir}(B^-\to\phi K^-)$, which motivates us to consider the second scenario where both left- and right-handed $s-s-Z^{\prime}$ couplings are considered. With the branching ratios and direct CP asymmetries of $B\to\phi K$ decays as constraints, the allowed regions for the $Z^{\prime}$ couplings are shown in blue in Fig.~\ref{ZpSpac4}. The pink regions shown in Fig.~\ref{ZpSpac4} are obtained with the bound $\phi_s^L=-91^{\circ}\pm33^{\circ}$, which is fitted under the constraints from $B\to \pi K$, $\pi K^{\ast}$ and $\rho K$ decays in scenario~I~(see Table~\ref{ZpCoupValue1}).

Before presenting our numerical results, we would like to firstly discuss the universality of the flavour-conserving $Z^{\prime}$ couplings for the first two generations. For the down-type quarks, the $3\times3$ diagonal $Z^{\prime}$ chiral charge matrices $\epsilon^{\psi_X}$~($X=L,R$) in the gauge eigenstate basis and the corresponding unitary matrices $V_{\psi_X}$ can be written, respectively, as
\ba
\epsilon^{\psi_X}= \left(\begin{array}{ccc}
\epsilon^{\psi_X}_d  & 0 & 0\\
0 & \epsilon^{\psi_X}_s& 0 \\
0& 0  & \epsilon^{\psi_X}_b
\end{array}\right)\,,\qquad
V_{\psi_X}= \left(\begin{array}{ccc}
V_{11}^X & V_{12}^X& V_{13}^X \\
V_{21}^X & V_{22}^X& V_{23}^X \\
V_{31}^X & V_{32}^X& V_{33}^X
\end{array}\right)\,.
\ea
Using Eq.~(\ref{3}), one can get the corresponding off-diagonal matrix element of $Z^{\prime}$ coupling in the mass eigenstate basis,
\be\label{Bx}
B_{ds}^{X}=V_{11}^X V_{21}^{X \ast} \epsilon^{\psi_X}_d + V_{12}^X V_{22}^{X \ast} \epsilon^{\psi_X}_s+V_{13}^X V_{23}^{X \ast}  \epsilon^{\psi_X}_b\,.
\ee
Moreover, using the unitarity of the matrices $V_{\psi_X}$,
\be\label{Vx}
V_{11}^X V_{21}^{X \ast}+ V_{12}^X V_{22}^{X \ast} +V_{13}^X V_{23}^{X \ast} =0\,,
\ee
and in the limit of small fermion mixing, Eq.~(\ref{Bx}) can be further simplified to
\be\label{rela}
B_{ds}^{X}\simeq V_{11}^X V_{21}^{X \ast}( \epsilon^{\psi_X}_d- \epsilon^{\psi_X}_s)\,.
\ee

\begin{figure}[t]
\begin{center}
\includegraphics [width=7.5cm]{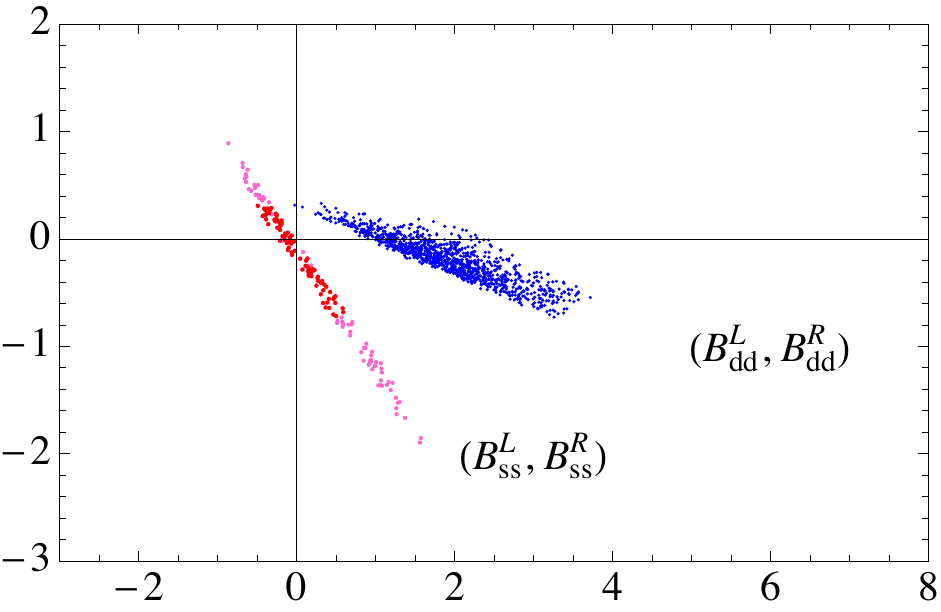}
\caption{\label{FCZpSpac}\small The allowed regions for the flavour-conserving $Z^{\prime}$ couplings $(B_{ss}^{L},B_{ss}^{R})$ and ($B_{dd}^{L},B_{dd}^{R})$ in the $B_{ii}^L-B_{ii}^R$ plane. The red region is obtained with the assumption that $B_{ss}^{R}=B_{dd}^{R}$.}
\end{center}
\end{figure}

It should be noted that, in this section, our studies of the $Z^{\prime}$ effects are performed in the ``SM limit", {\it i.e.,} the right-handed $Z^{\prime}$ coupling matrix is diagonal and hence no new types of four-quark operators arise compared to the SM ones given in Eq.~(\ref{eq:eff}). This implies $B_{ds}^{R}=0$ and, as indicated by Eq.~(\ref{rela}), $\epsilon^{\psi_R}_d=\epsilon^{\psi_R}_s$, which results in $B_{ss}^{R}=B_{dd}^{R}$. In order to check if our fitted $Z^{\prime}$ coupling parameters satisfy such a relation, we re-plot in Fig.~\ref{FCZpSpac} the allowed parameter spaces of the flavour-conserving $Z^{\prime}$ couplings $(B_{dd}^{L},B_{dd}^{R})$~(in blue in Fig.~\ref{ZpSpac1}, from $B\to\pi K$, $\pi K^{\ast}$ and $\rho K$ decays) and $(B_{ss}^{L},B_{ss}^{R})$~(in pink in Fig.~\ref{ZpSpac4}, from $B\to\phi K$ decays). It can be seen that, due to the fact that the allowed range of $B_{ss}^{R}$ is larger than that of $B_{dd}^{R}$, the relation $B_{ss}^{R}=B_{dd}^{R}$ can be easily satisfied. With the constraint $B_{ss}^{R}=B_{dd}^{R}$ assumed, the allowed ranges of $(B_{ss}^{L},B_{ss}^{R})$ are further reduced as is shown in red in Figs.~\ref{ZpSpac4} and \ref{FCZpSpac}. It is interesting to note that, by fitting the points in Fig.~\ref{FCZpSpac}, one may find an approximate linear relation between $B_{ss}^{L}$ and $B_{ss}^{R}$,
\be\label{relas}
B_{ss}^{L}\simeq-0.12-1.02B_{ss}^{R}\,,
\ee
which implies that the left-handed $Z^{\prime}$ coupling parameter $B_{ss}^{L}$ would also be further restricted.

However, as is shown in Fig.~\ref{FCZpSpac}, it is found that there is no overlap between the allowed regions of $(B_{dd}^{L},B_{dd}^{R})$ and $(B_{ss}^{L},B_{ss}^{R})$, which implies that $B_{dd}^{L}\neq B_{ss}^{L}$ and hence generally the off-diagonal element $B_{ds}^{L}\neq 0$. While $B_{ss}^{L}=B_{dd}^{L}$ and $B_{ds}^{L}=0$ are generally assumed due to the constraints from $K^0-\bar{K}^0$ mixing, a recent explicit investigation performed in Refs.~\cite{BurasZp,chiangK} indicate that the constraints from $K^0-\bar{K}^0$ mixing on the module of $B_{ds}^{L}$ are quite weak, because the observables $\triangle M_K$ and $\epsilon_K$ are governed, respectively, by the real and the imaginary part of $M^K_{12}$; it was found that the $Z^{\prime}$ contribution to $\epsilon_K$ even vanishes when $\phi_{sd}=n\pi/2$~\cite{BurasZp}. Thus, a nonzero $B_{ds}^{L}$ is at least not excluded under the constraints from $K^0-\bar{K}^0$ mixing, and hence the case with $B_{ss}^{L}\neq B_{dd}^{L}$ is still allowed.

Finally, our numerical results of the $Z^{\prime}$ couplings in scenario~I are presented in the second column of Table~\ref{ZpPareValueSS}. It is found that, while the ranges of $B_{ss}^{L,R}$ are severely restricted, their signs are hardly determined due to the interference effects between them. Furthermore, as is determined by Eq.~(\ref{relas}) and shown in Fig.~\ref{FCZpSpac}, the two parameters $B_{ss}^{L}$ and $B_{ss}^R$ could not simultaneously take the same signs, which is mainly required by the observable $A_{CP}^{dir}(B^-\to\phi K^-)$. With these numerical results of $Z^{\prime}$ couplings as inputs, our theoretical predictions for the observables of $B\to\phi K$ decays are then presented in the fourth column of Table~\ref{phiKobs}, from which one can find that the predicted $A_{CP}^{dir}(B^-\to\phi K^-)$ agrees well with the data.

\section{Conclusions}

In this paper, motivated by the latest experimental data of $B_s-\bar{B}_s$ mixing and various hadronic $b\to s$ transitions, we have performed a comprehensive reanalysis of the impact of a family non-universal $Z^{\prime}$ boson on these processes. For hadronic B-meson decays, our studies of the $Z^{\prime}$ effects are performed in the ``SM limit", {\it i.e.,} the right-handed $Z^{\prime}$ coupling matrix is diagonal and hence no new types of four-quark operators arise compared to the SM ones. Our main conclusions are summarized as follows:

\begin{itemize}
\item Among the several observables of $B_s-\bar{B}_s$ mixing, the precise $\Delta M_s$ and $\phi_s^{c\bar{c}s}$ put stringent constraints on the $Z^{\prime}$ coupling $B_{sb}^{L}$, whereas the constraints from $\Delta \Gamma_s$ and $a_{sl}^{s}$ are very weak due to the large experimental uncertainties. While the moduli $|B_{sb}^{L,R}|$ are stringently bounded, the weak phases $\phi_s^{L,R}$ are still not restricted by these updated experimental data. Numerically, we get $|B_{sb}^{L}|\leqslant 0.98\times10^{-3}$ and $|B_{sb}^{L,R}|\leqslant 0.43\times10^{-3}$ in scenarios LL and LR, respectively. Moreover, with $\phi_s^{L}=-91^{\circ}\pm33^{\circ}$, which is required to resolve the observed ``$\pi K$ CP puzzle", we get $|B_{sb}^{L}|\leqslant 0.83\times10^{-3}$.

\item The allowed parameter spaces of $Z^{\prime}$ couplings are found to satisfy the constraints from $B\to\pi K$, $\pi K^{\ast}$ and $\rho K$ decays, and hence could provide a possible solution to the observed ``$\pi K$ CP puzzle" through a sizable correction to the EW-penguin coefficient $\alpha^p_{3,{\rm EW}}(PP)=a_{9}^{p}-a_7^p$. Furthermore, the direct CP asymmetries of these hadronic B-meson decays put stringent constraints on the weak phase $\phi_s^{L}$ and the flavor-conserving $Z^{\prime}$ couplings. Our evaluations are performed in two different scenarios, with the corresponding numerical results of the $Z^{\prime}$ coupling parameters summarized in Tables~\ref{ZpCoupValue1} and~\ref{ZpCoupValue2}, respectively.

\item The $B_s\to K K$, $K K^{\ast}$ and $\pi^0 \phi$ decays, being induced by the same quark-level $b\to s q\bar{q}~(q=u,d)$ transitions, could provide further tests of such a family non-universal $Z^{\prime}$ model with the successful running of LHCb. Especially, as the $\bar{B}_s \to \pi^0 \phi$ decay is dominated by the EW-penguin coefficient $\alpha_{3,{\rm EW}}^p$ and is relatively ``clean", it would play a key role in revealing the observed ``$\pi K$ CP puzzle" and probing the proposed NP explanations.

\item To get information about the $s-s-Z^{\prime}$ coupling and check if the couplings of $Z^{\prime}$ boson to quarks are universal for the first two generations, we have also studied the penguin-dominated $B\to \phi K$ decays. The numerical results for $s-s-Z^{\prime}$ couplings are summarized in Table~\ref{ZpPareValueSS}.  It is found that, due to the large $A_{CP}^{dir}(B^-\to\phi K^-)$ reported by the BaBar collaboration, a significant $Z^{\prime}$ correction is required and the left-handed $s-s-Z^{\prime}$ coupling is different from the $d-d-Z^{\prime}$ one. However, as the LHCb measurement conflicts slightly with the BaBar data and their experimental uncertainties are still quite large, the refined measurements are required to either confirm or refute such a finding.
\end{itemize}

As a final comment, we would like to point out that, given the current lower limits on $m_{Z^{\prime}}\geq2~{\rm TeV}$ set by the ATLAS and CMS collaborations~\cite{ATLAS-Exotics,CMS-Exotics,Aad:2014cka,Chatrchyan:2012oaa}, the $Z^{\prime}$ effects on $B_s-\bar{B}_s$ mixing and hadronic B-meson decays are going to be small, and are usually plagued by large experimental and theoretical uncertainties. It is therefore quite difficult to deduce definitely the presence of $Z^{\prime}$ effects at the moment. With the running LHCb and the upcoming Super-KEKB experiments, together with the improved theoretical predictions, B physics is expected to enter a precision era, which would exhibit the exact features and the flavour structures of various NP models, including the $Z^{\prime}$ model considered in this paper.

\section*{Acknowledgments}
The work was supported by the National Natural Science Foundation of China~(NSFC) under Grant Nos.~11105043, 11005032, 11225523 and 11224504, Research Fund for the Doctoral Program of Higher Education of China under Grant Nos. 20114104120002 and 20104104120001. Q. Chang was also supported by a Foundation for the Author of National Excellent Doctoral Dissertation of P. R. China~(Grant No. 201317) and Program for Science and Technology Innovation Talents in Universities of Henan Province~(Grant No. 14HASTIT036). X.~Q. Li was also supported in part by the Scientific Research Foundation for the Returned Overseas Chinese Scholars, State Education Ministry.

\begin{appendix}

\section*{Appendix: Theoretical input parameters}

For the CKM matrix elements, we adopt the results~\cite{PDG12}
\begin{eqnarray}
&& |V_{us}|=0.2252\pm0.0009\,,\qquad
|V_{ub}|=0.00415\pm0.00049\,,\nonumber\\
&& |V_{cb}|=0.0409\pm0.0011\,,\qquad
\gamma=(68^{+10}_{-11})^{\circ}\,,
\end{eqnarray}
which are all extracted from tree-dominated processes and are, therefore, almost insensitive to physics beyond the SM.

As for the quark masses, we take~\cite{PDG12}
\begin{eqnarray}
 &&m_u=m_d=m_s=0, \qquad m_c=1.67\pm0.07\,{\rm GeV},\nonumber\\
 &&m_b=4.78\pm0.06\,{\rm GeV}, \qquad m_t=173.5\pm1.0\,{\rm GeV}\,,
\end{eqnarray}
for the pole masses and
\begin{eqnarray}
&& \frac{\overline{m}_s(\mu)}{\overline{m}_q(\mu)} = 27\pm1\,,\qquad
\overline{m}_{s}(2\,{\rm GeV})=95\pm5\,{\rm MeV}, \qquad
\overline{m}_{c}(\overline{m}_{c})=1.275\pm0.025\,{\rm GeV}\,\nonumber\\
&& \overline{m}_{b}(\overline{m}_{b}) = 4.18\pm0.03\,{\rm GeV}\,,\qquad
\overline{m}_{t}(\overline{m}_{t})=160.0^{+4.8}_{-4.3}\,{\rm GeV}\,,
\end{eqnarray}
for the running masses, where $m_q=m_{u,d}$. In addition, the value of mass parameter $m_b^{pow}=4.8^{+0.0}_{-0.2}$, which appears in the parameterization of the matrix elements $\langle B_s|O_i| \bar{B}_s \rangle $, is used.

The B-meson decay constants read~\cite{DecayCon}
\begin{flalign}
 f_{B_{s}}=(0.231\pm0.015)\,{\rm GeV}\,, \qquad  f_{B_{d}}=(0.190\pm0.013)\,{\rm GeV}\,,
\end{flalign}
and the ones of the other light mesons read
\begin{flalign}
& f_{\pi}=(130.4\pm0.2)\,{\rm MeV}\,, \qquad f_{K}=(156.1\pm0.8)\,{\rm MeV}\,,~\cite{PDG12}\\
& f_{K^{\ast}}=(217\pm5)~{\rm MeV}, \qquad f_{\rho}=(205\pm9)~{\rm MeV}\,, \qquad f_{\phi}=(215\pm5)~{\rm MeV}\,.~\cite{BallZwicky}
\end{flalign}
We take the following inputs for the heavy-to-light transition form factors~\cite{BallZwicky}
\begin{eqnarray}
 & &F^{B\to \pi}_{0}(0)=0.258\pm0.031\,, \qquad
     F^{B\to {K}}_{0}(0)=0.331\pm0.041\,,\nonumber\\
    &&A_0^{B\to K^\ast}(0)=0.374\pm0.034,  \qquad
    A_0^{B\to \rho}(0)=0.303\pm0.028,\nonumber\\
     &&A_0^{B_s\to K^\ast}(0)=0.360\pm0.034, \qquad
     A_0^{B_s\to \phi}(0)=0.474\pm0.033.
     \end{eqnarray}
For $\bar{B}_s\to K K$ decays, as is suggested in Ref.~\cite{Cheng2}, we shall use $F^{B_s\to K}_{0}(0)=0.24$ obtained by both lattice and pQCD calculations.

The B-parameters for $B_s-\bar{B}_s$ mixing read~\cite{BagPara}
\begin{flalign}
&B_1=0.86\pm0.02^{+0.05}_{-0.04}\,,\qquad B_2=0.83\pm0.02\pm0.04\,,\qquad B_3=1.03\pm0.04\pm0.09\,, \nonumber\\
&B_4=1.17\pm0.02^{+0.05}_{-0.07}\,,\qquad B_5=1.94\pm0.03^{+0.23}_{-0.07}\,.
\end{flalign}

\end{appendix}

 \end{document}